\documentclass[twocolumn,a4paper,reprint,amssymb,aps,prd,groupedaddress,superscriptaddress,amsmath,nobibnotes,showpacs,nofootinbib]{revtex4-2}

\usepackage{graphicx,bm,color,psfrag}
\usepackage{amsfonts}
\usepackage{lipsum}
\usepackage{float}
\usepackage{enumitem}
\usepackage[colorlinks = true,
            linkcolor = blue,
            urlcolor  = blue,
            citecolor = blue,
            anchorcolor = blue]{hyperref}
\usepackage[normalem]{ulem}
\usepackage{xcolor}

\usepackage{threeparttable,booktabs}

\usepackage{eurosym}
\usepackage{mathtools}
\usepackage{physics}

\begin{document}

\title{Equivalence of matter-type modified gravity theories to general relativity \\ with nonminimal matter interaction}

\author{\"{O}zg\"{u}r Akarsu}
\email{akarsuo@itu.edu.tr}
\affiliation{Department of Physics, Istanbul Technical University, Maslak 34469 Istanbul, Turkey}

\author{Mariam Bouhmadi-L\'opez}
\email{mariam.bouhmadi@ehu.eus}
\affiliation{IKERBASQUE, Basque Foundation for Science, 48011, Bilbao, Spain.}
\affiliation{Department of Physics and EHU Quantum Center, University of the Basque Country  UPV/EHU, P.O. Box 644, 48080 Bilbao, Spain.}
 
\author{Nihan Kat{\i}rc{\i}}
\email{nkatirci@dogus.edu.tr}
\affiliation{Department of Electrical and Electronics Engineering Do\u gu\c s University \"Umraniye, 34775 Istanbul, Turkey}

\author{Elham Nazari}
\email{elham.nazari@mail.um.ac.ir}
\affiliation{Department of Physics, Faculty of Science, Ferdowsi University of Mashhad, P.O. Box 1436, Mashhad, Iran}

\author{Mahmood Roshan}
\email{mroshan@um.ac.ir}
\affiliation{Department of Physics, Faculty of Science,
Ferdowsi University of Mashhad,
P.O. Box 1436, Mashhad, Iran}
\affiliation{School of Astronomy,
Institute for Research in Fundamental Sciences (IPM),
P. O. Box 19395-5531, Tehran, Iran}

\author{N. Merve Uzun}
\email{uzunmer@itu.edu.tr}
\affiliation{Department of Physics, Istanbul Technical University, Maslak 34469 Istanbul, Turkey}
\affiliation{Department of Physics, Bo\u{g}azi\c{c}i University, Bebek 34342 Istanbul, Turkey}

\begin{abstract}
In this study, we first establish that gravity models incorporating matter-related terms, such as $f(\mathcal{L}_{\rm m})$, $f(g_{\mu\nu} T^{\mu\nu})$, and $f(T_{\mu\nu} T^{\mu\nu})$, into the usual matter Lagrangian density $\mathcal{L}_{\rm m}$, are equivalent to general relativity (GR) with nonminimal matter interactions. Through the redefinition $\mathcal{L}_{\rm m}+f \rightarrow \mathcal{L}_{\rm m}^{\rm tot}$, these models are exactly GR, yet the usual material field $T_{\mu\nu}$ and its accompanying partner, namely, the modification field $T_{\mu\nu}^{\rm mod}$, engage in nonminimal interactions. Specifically, $\nabla^{\mu}T_{\mu\nu}=-Q_{\nu}=-\nabla^{\mu}T_{\mu\nu}^{\rm mod}$, where $Q_{\nu}$ is the interaction kernel that governs the rate of energy transfer. Our focus narrows on the specific model of $f(T_{\mu\nu} T^{\mu\nu})$, known as Energy-Momentum Squared Gravity (EMSG), where the usual material field $T_{\mu\nu}$ is accompanied by an \textit{energy-momentum squared field} (EMSF), $T_{\mu\nu}^{\rm emsf}$, along with a sui generis nonminimal interaction between them. We demonstrate that a particular $T_{\mu\nu}^{\rm emsf}$ can be introduced by \textit{removing} $\frac{\partial^2 \mathcal{L}_{\rm m}}{\partial g^{\mu\nu} \partial g^{\sigma\epsilon}}$ (the new term emerging in models that incorporate scalars formed from $T_{\mu\nu}$), thanks to the freedom in determining the interaction kernel, but this approach compromises the Lagrangian formulation of EMSG. Additionally, we address the ambiguities regarding the perfect fluid stemming from this new term. We show the proper way of calculating this term for a perfect fluid, revealing that it is indeed non-zero, contrary to common assumption in the literature. Finally, we re-examine cosmological models within the realm of EMSG, offering new insights into the applicability and interpretation of our findings in EMSG and similar theoretical frameworks.

\end{abstract}

\maketitle

\section{Introduction} 
Einstein’s general theory of relativity (GR) is in agreement with all the local tests to a precision of $10^{-5}$~\cite{Will:2014kxa} whereas the standard model of cosmology based on GR, Lambda cold dark matter ($\Lambda$CDM), suffers from a number of theoretical problems relevant to the cosmological constant $\Lambda$~\cite{Weinberg:1988cp,Peebles:2002gy,Padmanabhan:2002ji} as well as tensions between the observational constraints obtained from different data sets~\cite{DiValentino:2020vhf,DiValentino:2020zio,DiValentino:2020vvd,DiValentino:2020srs,DiValentino:2021izs,Perivolaropoulos:2021jda,Abdalla:2022yfr}. One main issue is the lack of explanation of cosmic dark sector, which consists of almost 95\% of the total energy budget of the present-day Universe, without invoking of some new material stresses such as scalar field. Also, the most statistically significant tension is in the estimation of the present-day value of the Hubble parameter $H_0$, i.e., the value measured from the cosmic microwave background (CMB) data by the Planck collaboration~\cite{Planck:2018vyg} is in about 5.0$\sigma$ tension with the model-independent local value reported by SH0ES collaboration~\cite{Riess:2021jrx}. Obviously, any modified theory of gravity that aims to bring resolutions to the issues in GR should not spoil the successful explanations of the Solar system phenomena such as the deflection of light, Shapiro time delay~\cite{Bertotti:2003rm,Shapiro:2004zz}, and the perihelion shift of Mercury~\cite{Mecheri2004NewVO,Antia:2007hm,Williams:2004qba,Williams:2005rv}, but manifests itself at cosmological scales (e.g., in the late universe relevant to the current accelerated expansion or in the early universe relevant to inflation, spatial anisotropy domination, initial singularity) and in the strong field regime (e.g., near/inside compact astrophysical objects such as neutron stars and black holes). 

Even though the vast majority of the modifications to GR focus on generalizing the gravitational Lagrangian density away from the linear function of scalar curvature ($R=g^{\mu\nu} R_{\mu\nu}$ with $R_{\mu\nu}$ being the Ricci curvature tensor, responsible for the Einstein tensor, $G_{\mu\nu}$, in Einstein’s field equations (EFE)~\cite{Copeland:2006wr,Caldwell:2009ix,Clifton:2011jh,Sotiriou:2008rp,DeFelice:2010aj,Capozziello:2011et,Nojiri:2017ncd,Nojiri:2010wj}) in the Einstein-Hilbert (EH) action, a new avenue  has been recently opened with modifying the introduction of the material source in the usual EH action. Such matter-type modified theories of gravity are constructed by either generalizing the matter Lagrangian density away from the linear function of $\mathcal{L}_{\rm m}$, viz., extending it to $f(\mathcal{L}_{\rm m})$~\cite{Harko:2010mv} or adding some analytic function of a scalar formed from the usual energy-momentum tensor (EMT), $T_{\mu\nu}$, of material stresses such as $f(g_{\mu\nu} T^{\mu\nu})$~\cite{Harko:2011kv} and $f(T_{\mu\nu} T^{\mu\nu})$~\cite{Katirci:2014sti,Roshan:2016mbt,Akarsu:2017ohj,Board:2017ign} into the EH action of GR. The first derivative of $T_{\mu\nu}$ with respect to the inverse metric tensor, $g^{\mu\nu}$, and particularly, the second metric derivative of $\mathcal{L}_{\rm m}$, viz., $\frac{\partial^2 \mathcal{L}_{\rm m}}{\partial g^{\mu\nu} \partial g^{\sigma\epsilon}}$, is a new term emerging in modified theories that contain a scalar formed from the usual EMT in their actions~\cite{Harko:2011kv,Haghani:2013oma,Odintsov:2013iba,Katirci:2014sti,Roshan:2016mbt,Akarsu:2017ohj,Board:2017ign,Asimakis:2022jel}. The absence of this term in earlier theories had required it to be handled carefully, however, it has been \textit{assumed} to be zero for perfect fluids in the literature to date without referring to a concrete explanation. Interestingly, the process of taking a closer look at the calculation of this term in the context of a particular theory dubbed \textit{energy-momentum squared gravity} (EMSG)~\cite{Katirci:2014sti,Roshan:2016mbt,Akarsu:2017ohj,Board:2017ign} has directed us to the fact that all the theories modifying the introduction of the material source in the usual EH action by adding only matter-related terms to $\mathcal{L}_{\rm m}$, in fact, cannot be considered as new theories of gravity but are minimal/nonminimal interaction models in the framework of GR. In other words, in this type of modifications, one basically deals with interacting two-field model correspondence in which the usual field, viz., $T_{\mu\nu}$, is accompanied by the partner field, $T_{\mu\nu}^{\rm mod}$, which is determined by the model under consideration. The interaction between the usual field and its partner can be rarely minimal, if combination of modification form and fluid type satisfy the local conservation, e.g., quadratic EMSG with dust. Or, generally both $T_{\mu\nu}$ and $T_{\mu\nu}^{\rm mod}$ violate the local conservation law due to nonminimal interaction between the usual field and its partner and we can treat these models as the nonminimal interaction models in the framework of GR. The matter-type modifications work properly at the level of field equations where the term $\frac{\partial^2 \mathcal{L}_{\rm m}}{\partial g^{\mu\nu} \partial g^{\sigma\epsilon}}$ is assumed as zero, which turns out to be a freedom that gives rise to a particular form of nonminimal interaction. However, we realize that these field equations are not actually derived from the actions they are considered to be derived when the second metric derivative of $\mathcal{L}_{\rm m}$ is not included in the interaction. Namely, contrary to the postulation that has been employed in the literature, as we will show in this paper, this term is not zero in general, but vanishes in particular cases (e.g., for canonical scalar fields) only~\cite{HosseiniMansoori:2023zop}, and thereby we are obliged to compromise the action/Lagrangian formulation of matter-type modifications that omit this term and confine ourselves to the field equations arising from these modifications.\footnote{This case is not limited to the EH action only; generalizations like $f(R,T)$, $f(R,T_{\mu\nu}T^{\mu\nu})$ [see Refs.~\cite{Sotiriou:2008rp,DeFelice:2010aj} for $f(R)$ theories], teleparallel gravity theories, viz. $f(\mathcal{T})$, $f(\mathcal{Q})$ [$\mathcal{T}$, $\mathcal{Q}$ are torsion and nonmetricity respectively, see  Refs.~\cite{Bahamonde:2021gfp,Hohmann:2021ast} for this class of theories] unified with the aforementioned modifications, and  models comprising nonminimal matter-curvature couplings such as $f(R_{\mu\nu}T^{\mu\nu})$~\cite{Haghani:2013oma,Odintsov:2013iba} and $f(G_{\mu\nu} T^{\mu\nu})$~\cite{Asimakis:2022jel} also suffer from the same problem relating to the actions/Lagrangians that describe them.} We would like to mention that one might utilize the correct expression for the second metric derivative of $\mathcal{L}_{\rm m}$ for perfect fluid, and in this way, derive the field equations in their full form starting from the actions proposed in studies on matter-type modifications to date. In the presence of a barotropic perfect fluid, the proper calculation of this term requires employing the equation of state (EoS) and sound speed (viz., $c_{\rm s}^{2}$) of the fluid, which we elaborate on in Section~\ref{sec:secderlm}. Nevertheless, the resulting model is still equivalent to a nonminimal interaction in GR, yet has a viable Lagrangian description.

  The EMSG theory was constructed by adding the term $f(T_{\mu\nu} T^{\mu\nu})$, envisaged as a correction, to the standard matter Lagrangian density, $\mathcal{L}_{\rm m}$, in the EH action. Therefore, as we switch to the framework of nonminimally interacting two-field model, the usual material field, $T_{\mu\nu}$, described by $\mathcal{L}_{\rm m}$ is accompanied by a new material field, $T_{\mu\nu}^{\rm emsf}$, which we call \textit{energy-momentum squared field} (EMSF) described by an arbitrary function of the scalar $T_{\mu\nu} T^{\mu\nu}$ formed solely from the usual field itself. Both material fields violate the local energy-momentum conservation law due to the nonminimal nature of the interaction between them, but in total behave like a single conserved material field (i.e., $T_{\mu\nu}^{\rm tot}$) having a vanishing divergence, which reveals that the theory under consideration is indeed GR.  In the following, we mention some recent studies of EMSG supporting our arguments and comment on them from the EMSF interaction perspective. For instance, it has been shown in Ref.~\cite{Akarsu:2022abd} that the interaction model in the presence of \textit{energy-momentum powered field} (EMPF), a specific form of EMSF described by the choice of the function $f(T_{\mu\nu} T^{\mu\nu})=\alpha (T_{\mu\nu} T^{\mu\nu})^{\eta}$ with $\alpha$ and $\eta$ being constants, leads to the gravitational potential form, post-parameterized Newtonian parameters, and geodesics for the test particles of GR. The parameter $\alpha$ determines the amount of EMSF with respect to the usual field and its dimension depends on the power, $\eta$.\footnote{Depending on the sign of the parameter $\alpha$, ghost/gradient instabilities may arise. For further details on this phenomenon, see, e.g., Ref.~\cite{HosseiniMansoori:2023zop}.} As expected, for a proper analysis, the slow motion condition is required to be described by making use of the matter variables of $T_{\mu\nu}^{\rm tot}$ in which the interaction is embedded since it combines the usual material field with its accompanying EMSF. However, the only difference is that for GR in the presence of minimally interacting material fields, we simply have the relation $M_{\rm ast}=M$, which means the mass of an astrophysical object inferred from astronomical observations (from planetary orbits, deflection of light, etc.) is equal to the actual physical mass whereas in GR with the EMPF interaction, we have $M_{\rm ast}=M+M_{\rm empf}$, that is, the same astrophysical mass corresponds to the sum of the physical mass and the contribution due to EMPF. As we take into consideration the fact that $M_{\rm empf}=M_{\rm empf}(\alpha,\eta,M)$, it is possible to infer not only $M_{\rm ast}$ alone from astronomical observations, but $M$ and $M_{\rm empf}$ separately if there is any information about the model parameters ($\alpha, \eta$) or $M$ from other independent phenomena such as cosmological observations and the structure of the astrophysical object. 
Moreover, in Ref.~\cite{Nazari:2022fbn}, considering the neutron stars as a deflector and knowing the density of this compact astrophysical object, the motion of light in the weak-field limit of \textit{quadratic EMSF}, corresponding to the $\eta=1$ case of EMPF, has been studied. It has been shown that the overall behavior of the light curves in this model matches those of GR with minimal interaction. Furthermore, the gravitational radiation properties of the quadratic EMSF as well as of \textit{scale-independent EMSF} (the case $\eta=\frac{1}{2}$ in EMPF) have been investigated using the post-Newtonian approximation in Refs.~\cite{Nazari:2022xhv,Akarsu:2023qnr}. It has been shown that in the case of binary systems as the source of gravitational waves, these interaction models only change the gravitational wave amplitude and not the wave polarizations, at least up to the post-Newtonian order considered in these studies. However, these effects on the wave amplitude do not mean departure from GR since they can be absorbed in the chirp mass definition of binaries.

A number of works in various contexts from cosmology to astrophysics demonstrate that EMSF interaction manifests itself at both high energy densities (viz., in the early Universe and inside the compact objects) and low energy densities (viz., in the late Universe)~\cite{Katirci:2014sti,Roshan:2016mbt,Akarsu:2017ohj,Board:2017ign,Akarsu:2018zxl,Nari:2018aqs,Akarsu:2018aro,Akarsu:2019ygx,Faria:2019ejh,Bahamonde:2019urw,Chen:2019dip,Barbar:2019rfn,Kazemi:2020hep,Singh:2020bdv,Nazari:2020gnu,Rudra:2020rhs,Akarsu:2020vii,Chen:2021cts,Tangphati:2022acb,Khodadi:2022xtl,Acquaviva:2022bju,Khodadi:2022zyz,2023arXiv230401571F}.  Here, we outline several promising features and capabilities from a cosmological perspective: The usual EMT violates the local conservation in general~\cite{Akarsu:2017ohj,Board:2017ign}, and it is possible to drive late time acceleration from the usual cosmological sources without invoking a cosmological constant $\Lambda$~\cite{Akarsu:2017ohj}. A source with constant inertial mass density arises in the interaction with \textit{energy-momentum log field} (EMLF) described by $f(T_{\mu\nu} T^{\mu\nu})=\alpha \ln{( \lambda T_{\mu\nu} T^{\mu\nu})}$ with $\lambda$ being a constant, and accordingly, it provides us with screening of $\Lambda$ in the past by altering the contribution of dust to the Friedmann equation~\cite{Akarsu:2019ygx}; dust interacting with its accompanying quadratic EMSF is able to screen the shear scalar (viz., the contribution of the expansion anisotropy to the average expansion rate of the universe), and thereby can lead to exactly the same Friedmann equation of the standard $\Lambda$CDM model even in the presence of anisotropic expansion~\cite{Akarsu:2020vii}. Also, EMSF interaction alters the cosmic history of the Universe by affecting the past or far future depending on the chosen model~\cite{Roshan:2016mbt,Akarsu:2018zxl,Acquaviva:2022bju}. On the astrophysical side, we expect that some distinguishing deviations can be achieved from the physical systems which have charge distribution. For instance, in Ref.~\cite{Roshan:2016mbt}, the charged black hole solution of the interaction with quadratic EMSF has been derived, and is different from the standard Reissner-Nordström spacetime metric. Also, EMSF does not lead to a nonsingular electric field for a point charge. In Ref.~\cite{Chen:2019dip}, the axial perturbations of the charged black holes have been studied for the nonminimal EMSF interaction. Unlike the minimal interaction, the correspondence between the eikonal quasinormal modes and the photon rings of the blackholes is broken~\cite{Chen:2019dip,Chen:2021cts} and this violation can be an important tool to test interaction models (kernels) via the upcoming gravitational wave observations.\footnote{For Maxwell field, in studies \cite{Roshan:2016mbt,Chen:2019dip,Chen:2021cts}, authors include the second metric variation of the matter Lagrangian density in their models whereas in the current study, we give a common recipe for EMSF by removing this term also for material fields other than perfect fluid, even if it exists.} Besides all these, following the interaction perspective first declared in the current study, Ref.~\cite{Akarsu:2023agp} has explored the implications of quadratic EMSF in the presence of additional relativistic relics and showcased the model's potential to accommodate deviations from standard cosmology and the Standard Model of particle physics via the most stringent cosmological bounds on $\alpha$.

This paper consists of three main parts: In Section~\ref{sec:nonmin}, we first demonstrate why the models that modify the usual EH action by adding only matter-related terms do not give rise to new theories of gravity, but instead correspond to GR with nonminimal interactions; we extensively discuss this new interpretation of such models, known as matter-type modified theories of gravity, and present its applications and implications by analyzing a case study: EMSG. In Section~\ref{sec:pf-EMT}, to lay the groundwork for subsequent calculations and interpretations, we revisit the derivation of the first metric derivatives of the matter Lagrangian densities of a perfect fluid, $\mathcal{L}_{\rm m}=p$ and $\mathcal{L}_{\rm m}=- \rho$, from thermodynamics, and hence, how its usual EMT and equations of motion are obtained via the variational method in GR. Then, we present the proper way of calculating the second metric derivatives of both matter Lagrangian densities.  In Section~\ref{emsg-cosm}, we reconsider the cosmological models in the framework of EMSG from the perspective of EMSF interaction. Finally, we conclude in the last section. To assist readers in navigating the paper, a flow scheme of the study is provided in Fig.~\ref{fig:scheme} in Appendix\ref{app:scheme}.

\section{Matter-type modifications as nonminimal interactions in General Relativity} \label{sec:nonmin}
  The EFE of GR, $G_{\mu\nu}= \kappa T_{\mu\nu}$, is derived from the EH action expressed by
\begin{equation}
 \mathcal{S}=\frac{1}{2 \kappa}\int {\rm d}^4x\, \sqrt{-g} R + \mathcal{S}_{\rm m},
\end{equation}
where $\kappa=8 \pi G$ ($G$ being the Newton's constant), $g$ is the determinant of the metric tensor $g_{\mu\nu}$, viz., $g={\rm det} (g_{\mu\nu})$, and $R=g^{\mu\nu} R_{\mu\nu}$ is the scalar curvature with $R_{\mu\nu}$ being the Ricci curvature tensor. The variation of the curvature part of this action with respect to the inverse metric $g^{\mu\nu}$ gives rise to the Einstein tensor, $G_{\mu\nu}= R_{\mu\nu}-\frac{1}{2} R g_{\mu\nu}$. The matter part of the action is defined as 
\begin{equation} \label{matteraction}
\mathcal{S}_{\rm m}=\int {\rm d}^4x\, \sqrt{-g} \mathcal{L}_{\rm m},
\end{equation}
where $\mathcal{L}_{\rm m}$ is the matter Lagrangian density describing the material field. Thereby, in order to reach the EFE, the EMT of any material field is defined as
\begin{align}
  \label{gen-tmunudef}
 T_{\mu\nu}=-\frac{2}{\sqrt{-g}}\frac{\delta(\sqrt{-g}\mathcal{L}_{\rm m})}{\delta g^{\mu\nu}}.
 \end{align}
We will now investigate matter-type modifications to GR from a different point of view. The usual action of matter-type modified theories are described by 
\begin{equation}
S=\int {\rm d}^4x\,\sqrt{-g}\left[\frac{1}{2\kappa} R+f(\mathcal{L}_{\rm m}, g_{\mu\nu} T^{\mu\nu}, T_{\mu\nu} T^{\mu\nu})+\mathcal{L}_{\rm m}\right],   \end{equation}
where $f$ can be any analytic function of material-related scalars: $\mathcal{L}_{\rm m}$~\cite{Harko:2010mv}, $g_{\mu\nu} T^{\mu\nu}$~\cite{Harko:2011kv} and $T_{\mu\nu} T^{\mu\nu}$~\cite{Katirci:2014sti,Roshan:2016mbt,Akarsu:2017ohj,Board:2017ign}. However, since the definition given in Eq.~\eqref{gen-tmunudef} is arranged in such a way that $T_{\mu\nu}$ emerges on the right-hand side of EFE, the matter part of the action can be redefined as 
\begin{equation}\label{Lmtot}
\mathcal{L}_{\rm m}^{\rm tot}=\mathcal{L}_{\rm m}+f,
\end{equation}
and imitating the definition stated in Eq.~\eqref{gen-tmunudef} gives rise to the total EMT as follows:
\begin{align} \label{emt-tot}
T_{\mu\nu}^{\rm tot}=-\frac{2}{\sqrt{-g}}\frac{\delta(\sqrt{-g}\mathcal{L}_{\rm m}^{\rm tot})}{\delta g^{\mu\nu}}.
\end{align}
Here $T_{\mu\nu}^{\rm tot}=T_{\mu\nu}+T_{\mu\nu}^{\rm mod}$ where $T_{\mu\nu}$ is the usual material field defined by $\mathcal{L}_{\rm m}$ and $T_{\mu\nu}^{\rm mod}$ is the accompanying material field defined by the function $f$ which characterizes the modification. We emphasize that by changing the matter field portion of GR, i.e., $\mathcal{S}_{\rm m}\longrightarrow\mathcal{S}_{\rm m}^{\rm tot}$ or $\mathcal{L}_{\rm m}\longrightarrow \mathcal{L}_{\rm m}^{\rm tot}$, like the modification imposed in the matter-type modified theories [see Eq.~\eqref{Lmtot}], the variational method indicates that the EMT is modified, and this new EMT is indeed conserved, $\nabla_{\mu}T^{\mu\nu}_{\rm tot}=0$. However, due to the extra term $f$, the equations of motion of the fluid in the GR are not satisfied anymore. In the following, utilizing the twice contracted Bianchi identity at the level of field equations, we point out this fact again.

On the other hand, we recall that in the presence of two or more material stresses, the EFE of GR can be written as $G_{\mu\nu}=\kappa T_{\mu\nu}^{\rm tot}$, where $T_{\mu\nu}^{\rm tot}$ denotes the sum of the EMTs of different material fields. Regarding the matter-type modifications, we realize that the actions of these models are indeed the usual EH action,  and hence, $T_{\mu\nu}$ and $T_{\mu\nu}^{\rm mod}$behave like two nonminimally interacting material fields \footnote{At this point, it should be noted that the modification field is determined by only the terms belonging to the usual material field via the function $f$, and hence, we can not regard them as two \textit{independent} material fields. Namely, one first introduces $T_{\mu\nu}$, and then defines $T_{\mu\nu}^{\rm mod}$ over it so that both are subject to a single set of initial conditions. In this study, we opt to interpret the situation as nonminimal matter interaction between the usual field and its \textit{accompanying partner} field. However, since $T_{\mu\nu}^{\rm mod}$ does not introduce an extra degree of freedom in the field-theoretical sense, it could either correspond to self-interacting models of the usual field or be considered as a modification of the usual EMT. Which of these interpretations better describes it will be determined in due course by the research community working on these models.} coupled to the spacetime in accordance with GR since the field equations of these models can be recast as
\begin{equation}
G_{\mu\nu} = \kappa T_{\mu\nu}+\kappa T_{\mu\nu}^{\rm mod},
\label{modfieldeq}
\end{equation}
which along with the twice contracted Bianchi identity implies the conservation of the total EMT, i.e., $\nabla^{\mu} T_{\mu\nu}^{\rm tot}=0$.  
Therefore, we have 
\begin{equation} \label{separatediv}
\nabla^{\mu}(T_{\mu\nu}+T_{\mu\nu}^{\rm mod})=0,
\end{equation}
but not necessarily $
\nabla^{\mu}T_{\mu\nu}=0$ and $\nabla^{\mu} T_{\mu\nu}^{\rm mod}=0$. In other words, from Eq.~\eqref{separatediv}, a possible nonminimal interaction between these two EMTs can be expressed as follows;
\begin{equation}
\label{eq:int}
\nabla^{\mu}T_{\mu\nu}=-Q_{\nu} \quad \textnormal{and} \quad \nabla^{\mu}T_{\mu\nu}^{\rm mod}=Q_{\nu},
\end{equation}
where the four-vector $Q_{\nu}$ is the interaction kernel that governs the rate of energy transfer. We point out that GR does not impose any condition on the interaction kernel $Q_{\nu}$. Namely, taking $Q_{\nu}=0$ (minimal interaction) in the usual fashion is in fact one of the possible freedoms, and it is this choice that leads to the conservation of each EMT individually. However, in matter-type modifications, the interaction kernel, $Q_{\nu}$, is determined by a general analytic function of the matter Lagrangian density $f(\mathcal{L}_{\rm m})$~\cite{Harko:2010mv} or of scalars constructed from the usual material field, $T_{\mu\nu}$, such as $f(g_{\mu\nu} T^{\mu\nu})$~\cite{Harko:2011kv} and $f(T_{\mu\nu} T^{\mu\nu})$~\cite{Katirci:2014sti,Roshan:2016mbt,Akarsu:2017ohj,Board:2017ign}. In the following section, we will elaborate on this new interpretation of such models, called matter-type modified theories of gravity, by analyzing a case study: energy-momentum squared gravity (EMSG) described by $f(T_{\mu\nu}T^{\mu\nu})$.

\subsection{Energy-Momentum Squared Field (EMSF) }   \label{EMSF}
 To investigate the interacting two-field models in GR, one starts from the continuity equations for each field by assuming an arbitrary interaction kernel between them in the background. Contrarily, the interaction kernel due to EMSG has a covariant formulation and is not completely arbitrary, but determined by the usual material stresses via the arbitrary function of the Lorentz scalar $T_{\mu\nu}T^{\mu\nu}$, viz., $f(T_{\mu\nu}T^{\mu\nu})$. Since the matter fields usually couple only to the metric tensor, we assume that $\mathcal{L}_{\rm m}$ depends only on $g^{\mu\nu}$ and not on its derivatives, use the relation $\delta \sqrt{-g}=-\frac{1}{2} \sqrt{-g} g_{\mu\nu} \delta g^{\mu\nu}$, and hence, obtain 
 \begin{align}
  \label{tmunudef}
T_{\mu\nu}=\mathcal{L}_{\rm m} g_{\mu\nu}-2\frac{\partial \mathcal{L}_{\rm m}}{\partial g^{\mu\nu}},
 \end{align}
valid for the Maxwell field and gauge fields in general, as well as for scalar fields and perfect fluids---spinor fields whose matter Lagrangian densities depend on also derivatives of the metric need a different treatment, Einstein-Cartan(-Sciama-Kibble) theory of gravity, by reformulating general relativity in terms of tetrad fields, are thereby excluded in the current study, see Ref.~\cite{Leclerc:2005na} for details. Note that this assumption allows us to use the metric variations and metric derivatives interchangeably. Inspired by the usual definition of the EMT given in Eq.~\eqref{gen-tmunudef}, in order to determine the form of the EMSF, we begin with the following expression  
 \begin{align}
  \label{emsgdef}
 T_{\mu\nu}^{\rm mod}=-\frac{2}{\sqrt{-g}}\frac{\delta(\sqrt{-g}f)}{\delta g^{\mu\nu}}=f\,g_{\mu\nu}-2f_{\mathbf{T^2}}\theta_{\mu\nu},
 \end{align}
 where 
 \begin{equation}
 \begin{aligned}
 \label{thetadef}
 f_{\mathbf{T^2}} \equiv \frac{\partial f}{\partial(T_{\sigma \epsilon} T^{\sigma \epsilon})} \quad ,\quad  \theta_{\mu\nu}\equiv \frac{\delta (T_{\sigma\epsilon}T^{\sigma\epsilon})}{\delta g^{\mu\nu}}.
 \end{aligned}
  \end{equation}
Consequently, substituting   Eq.~\eqref{emsgdef} into Eq.~\eqref{modfieldeq}, the field equations become
\begin{equation}
G_{\mu\nu} = \kappa T_{\mu\nu}+\kappa\left(f g_{\mu\nu}-2 f_{\mathbf{T^2}} \theta_{\mu\nu}\right).
\label{EMSGfieldeq}
\end{equation}
Therefore, in this model,  the corresponding interaction kernel defined in Eq.~\eqref{eq:int} has a particular form as 
\begin{equation}
\label{eq:int1}
  Q_{\nu}=\nabla_{\nu}f-2 \,\theta_{\mu\nu} \nabla^{\mu} f_{\mathbf{T^2}} -2f_{\mathbf{T^2}}\nabla^{\mu} \theta_{\mu\nu}, 
\end{equation}
due to the function $f$ and the tensor $\theta_{\mu\nu}$ governing the modification field, $T_{\mu\nu}^{\rm mod}$. One can immediately notice that even simple choices for the $f$ function can lead to nontrivial interaction kernels with covariant formulation.

Comparing to the common interacting two-fluid models in the literature, we have two degrees of freedom in constructing the interaction kernel for the usual material field and the  modification field. For instance, in a two-component perfect fluid in the Friedmann-Lemaitre-Robertson-Walker (FLRW) universe, one has five unknowns ($\rho_1, p_1, \rho_2, p_2, a$) and two Friedmann equations with $a$ being the scale factor. The remaining three degrees of freedom correspond to the two equations of state that characterize the fluids and the form of the interaction between these fluids.\footnote{In the context of astrophysics, there are nine unknowns: the rest-mass density, total energy density, and pressure of the fluid 1 ($\rho_1^0, \rho_1, p_1$), those of the fluid 2 ($\rho_2^0, \rho_2, p_2$), and three components of $u^{\mu}$. Here, $\rho^0$ stands for the rest-mass density of the fluid. One has four equations from the conservation of EMT and one equation from the conservation of rest-mass density. Hence, four degrees of freedom are the EoS of fluid 1, the EoS of fluid 2, the form of the interaction in EMT conservation, and the form of the interaction in rest-mass conservation. In scalar field case, equations of state are determined by two potentials $V_1(\phi_1)$ and $V_2(\phi_2)$ corresponding to two scalar fields $\phi_1$ and $\phi_2$. However, Maxwell field yields an anisotropic-pressure source with $p_{x}=-p_{ y}=-p_{z}$, and in this case two-field model is not compatible with the FLRW spacetime metric.} Therefore, the presence of the function $f$ and the new tensor $\theta_{\mu\nu}$ in both Eqs.~\eqref{emsgdef} and~\eqref{eq:int1} implies that the EoS of the second source described by $ T_{\mu\nu}^{\rm mod}$ and the interaction kernel are intertwined in this model. Hence, employing only the function $f$ to construct the form of the interaction kernel, the missing EoS of the source generated by $ T_{\mu\nu}^{\rm mod}$ provides us with the freedom to adjust the expression for $\theta_{\mu\nu}$. So, we first determine the structure of $\theta_{\mu\nu}$ that can be used in common in the presence of material fields whose EMTs are calculated via Eq.~\eqref{tmunudef}, and then render only the function $f$ responsible for identifying the different forms of the interaction.   

From the definition given in Eq.~\eqref{tmunudef}, we find the following expression for $\theta_{\mu\nu}$ in terms of the EMT and the matter Lagrangian density of the usual material stress as follows:
\begin{equation}
\begin{aligned}  
\label{theta}
\theta_{\mu\nu}&=T^{\sigma\epsilon}\frac{\delta T_{\sigma\epsilon}}{\delta g^{\mu\nu}}+T_{\sigma\epsilon}\frac{\delta T^{\sigma\epsilon}}{\delta g^{\mu\nu}}\\
&=-2\mathcal{L}_{\rm m}\left(T_{\mu\nu}-\frac{1}{2}g_{\mu\nu}T\right)-T T_{\mu\nu}\\
&\quad\,+2T_{\mu}^{\lambda}T_{\nu\lambda}-4T^{\sigma\epsilon}\frac{\partial^2 \mathcal{L}_{\rm m}}{\partial g^{\mu\nu} \partial g^{\sigma\epsilon}},
\end{aligned}
\end{equation}
where $T=g_{\mu\nu} T^{\mu\nu}$ is the trace of the EMT. We should remark that the second variation of $\mathcal{L}_{\rm m}$ with respect to the inverse metric tensor emerges from the first variation of $T_{\mu\nu}$, and is a new term that did not exist in the literature before. Unfortunately, this last term in Eq.~\eqref{theta} has been considered to be zero for perfect fluids without providing a concrete proof in studies on EMSG~\cite{Katirci:2014sti,Roshan:2016mbt,Akarsu:2017ohj,Board:2017ign} as well as on other models that contain scalars formed from the usual EMT like $g_{\mu\nu} T^{\mu\nu}$~\cite{Harko:2011kv}, $R_{\mu\nu}T^{\mu\nu}$~\cite{Haghani:2013oma,Odintsov:2013iba}  and $G_{\mu\nu} T^{\mu\nu}$~\cite{Asimakis:2022jel}. However, if the matter Lagrangian density is chosen either $\mathcal{L}_{\rm m}=p$ or $\mathcal{L}_{\rm m}= - \rho$ for the perfect fluid case, this term includes the metric variations of both the energy density and pressure, which exist separately so far in the literature, cf. Eqs.~\eqref{varpGR} and~\eqref{varrhoGR} in Section~\ref{firstvar}. Here we only state that this term does not vanish in general and refer the reader to Section~\ref{sec:secderlm} in which we provide the detailed calculation. On the other hand, while determining the interaction kernel, we can \textit{omit} this term and avoid these variations emerging simultaneously thanks to the freedom we have. Therefore, we define the EMSF from Eq.~\eqref{emsgdef} as follows
\begin{align}
  \label{def-EMSF}
T_{\mu\nu}^{\rm emsf}=f\,g_{\mu\nu}-2f_{\mathbf{T^2}}\theta_{\mu\nu}^{\rm emsf},
\end{align}
but adhering to the convention in the literature \textit{remove} the second derivative of $\mathcal{L}_{\rm m}$ from Eq.~\eqref{theta} and arrive at the following new tensor for the EMSF
\begin{align} \label{theta-EMSF}
\theta_{\mu\nu}^{\rm emsf}=-2\mathcal{L}_{\rm m}\left(T_{\mu\nu}-\frac{1}{2}g_{\mu\nu}T\right) -T T_{\mu\nu}+2T_{\mu}^{\lambda}T_{\nu\lambda}.    \end{align}
We note that compared to Eq.~\eqref{theta}, Eq.~\eqref{theta-EMSF} does not equal to the metric variation of the Lorentz scalar $T_{\mu\nu} T^{\mu\nu}$ anymore, and hence, $T_{\mu\nu}^{\rm emsf}$ does not correspond to the metric variation of $\sqrt{-g} f$ in contrast to $T_{\mu\nu}^{\rm mod}$ given in Eq.~\eqref{emsgdef}. It is obvious that the aforementioned freedom of being able to remove the last term from Eq.~\eqref{theta} has prevented the models studied in the literature to date from any inconsistencies since this term is said to be assumed as zero though it is not indeed. Also, this choice allows us to properly describe a dust fluid for the case $\mathcal{L}_{\rm m}= p$ in this model because the proper calculation of the last term in Eq.~\eqref{theta} generates a divergence when $p=0$, cf. Eq.~\eqref{del2p} for the relevant expression. Note that even though this term is not problematic for other material stresses like scalar and Maxwell fields\footnote{For the canonical scalar field $\phi$ described by $\mathcal{L}_{\phi}^{\rm c}=X - V$ where $X=-\frac{1}{2}\nabla_{\alpha} \phi \nabla^{\alpha} \phi$ is the kinetic part and  $V=V(\phi)$ is the potential part, $\frac{\partial^2 \mathcal{L}_{\phi}^{\rm c}}{\partial g^{\mu \nu} \partial g^{\sigma\epsilon}}=0$~\cite{Chen:2019dip} whereas the noncanonical generalization of the scalar field described by $\mathcal{L}_{\phi}^{\rm nc}=F(X) - V$ satisfies $\frac{\partial^2 \mathcal{L}_{\phi}^{\rm nc}}{\partial g^{\mu \nu} \partial g^{\sigma\epsilon}}=\frac{1}{4} \frac{{\rm d}^2 F}{{\rm d} X^2} \nabla_{\mu} \phi \nabla_{\nu} \phi \nabla_{\sigma} \phi \nabla_{\epsilon} \phi $ with $F$ being an arbitrary function of $X$. For the Maxwell field described by $\mathcal{L}_{\rm m}=-\frac{1}{4} F_{\mu\nu}F^{\mu\nu}$, we have $\frac{\partial^2 \mathcal{L}_{\rm m}}{\partial g^{\mu\nu} \partial g^{\sigma\epsilon}}=-\frac{1}{2} F_{\sigma\mu}F_{\epsilon\nu}$ where $F_{\mu\nu}$ is the usual electromagnetic field strength tensor~\cite{Roshan:2016mbt}. 
}, for consistency, the definition~\eqref{def-EMSF} should be used for these fields as well.

Therefore, not only because of the freedom we have in determining  $\theta_{\mu\nu}$, but also to avoid any diverging term in this tensor, it is reasonable to omit the second metric variation of $\mathcal{L}_{\rm m}$ in accordance with the convention in the literature. However, this choice comes at a price: $\mathcal{L}_{\rm m}^{\rm tot}=\mathcal{L}_{\rm m}+f$ is not the total matter Lagrangian density of the model anymore when a term arising from the variation of $\sqrt{-g} f$ is omitted.
This point indicates that by removing the second derivative of $\mathcal{L}_{\rm m}$, the EMSF interaction is not derived from a well-defined Lagrangian density anymore, and thereby, we are obliged to compromise the Lagrangian formulation of this interaction model. We will momentarily set aside the discussion on nonminimal matter interactions to address issues related to perfect fluids. The proper calculations of the first and second metric derivatives of the matter Lagrangian density for perfect fluids, from a thermodynamic perspective, will be demonstrated in the following section.

\section{Revealing and Fixing the  Ambiguities About Perfect Fluid} \label{sec:pf-EMT}

Perfect fluids (often used to model idealized distributions of matter in settings ranging from compact astrophysical objects to cosmology) are described by the EMT of the form
\begin{align}
\label{em}
T_{\mu\nu}^{\rm pf}=(\rho+p)u_{\mu}u_{\nu}+p g_{\mu\nu},
\end{align} 
where $\rho>0$ and $p$ are, respectively, the fluid's energy density and thermodynamic pressure measured by an observer comoving with the fluid, $u^{\mu}$ is the fluid's four-velocity satisfying the condition $u_{\mu}u^{\mu}=-1$, and accordingly, $u_{\mu} \nabla_{\nu} u^{\mu}=0$. It is noteworthy that any material field whose EMT is of the above form, whether or not it is derived from a Lagrangian, is called a perfect fluid~\cite{Hawking:1973uf}. On the other hand, the definition of the matter Lagrangian density that gives rise to the perfect fluid EMT through the definition given in Eq.~\eqref{tmunudef} is not unique; either $\mathcal{L}_{\rm m}=p$ or $\mathcal{L}_{\rm m}=- \rho$ results in the same EMT, viz., $T^{\mu\nu}$ that describes perfect fluid matter distributions as given in Eq.~\eqref{em}. For details and other possible choices for the Lagrangian density of the perfect fluid, see Ref.~\cite{Brown:1993} and the references therein.

\subsection{The first metric derivative of the matter Lagrangian density for perfect fluid}
\label{firstvar}
At this point, it can be easily seen that the energy density and pressure depend on the metric tensor, and these dependencies separately provide us with the perfect fluid EMT~\eqref{em} through the definition given in~\eqref{tmunudef}. In other words, if we choose $\mathcal{L}_{\rm m}=p$ for the matter Lagrangian density of the perfect fluid, the definition in Eq.~\eqref{tmunudef} along with the perfect fluid EMT given in Eq.~\eqref{em} imply
\begin{align}
 \mathcal{L}_{\rm m}=p \quad\rightarrow\quad \frac{\delta p}{\delta g^{\mu\nu} }=-\frac{1}{2} (\rho+p)\,u_{\mu} u_{\nu}. \label{varpGR}
\end{align}
On the other hand, for the choice of $\mathcal{L}_{\rm m}=-\rho$ to obtain the same perfect fluid EMT~\eqref{em}, the definition~\eqref{tmunudef} requires
\begin{align}
\mathcal{L}_{\rm m}=- \rho \quad\rightarrow\quad \frac{\delta \rho}{\delta g^{\mu\nu} }=\frac{1}{2} (\rho+p) (u_{\mu} u_{\nu}+g_{\mu\nu})  \label{varrhoGR}.
\end{align}
However, notice that what is done here is merely a deduction in reverse order. In what follows, to indicate assumptions behind each case, we shall obtain the perfect fluid EMT from the matter Lagrangian densities $\mathcal{L}_{\rm m}=p$ and $\mathcal{L}_{\rm m}=-\rho$ by making use of the first law of thermodynamics. We will follow the procedure presented in Refs.~\cite{Taub:1954, Schutz1970,Brown:1993}.

\subsubsection{The matter Lagrangian density $\mathcal{L}_{\rm m}=p$}  \label{sec:Lm-p}
 We let the EoS be given as $p=p(h,s)$ and the matter Lagrangian density be defined as $\mathcal{L}_{\rm m}=p$. Here, $h= \frac{\rho+p}{n}$ is the specific enthalpy with $n$ being the particle number density and $s$ is the specific entropy, i.e., the entropy per unit mass. Then, we have $\mathcal{S}_{\rm m}=\int {\rm d}^4x\, \sqrt{-g}\, p $ for the action of the perfect fluid, and taking its variation, we obtain
\begin{align}
 \delta  \mathcal{S}_{\rm m}= \int {\rm d^4}x (p \, \delta \sqrt{-g}+\sqrt{-g} \delta p).
\end{align}
In order to derive the variation of $\mathcal{S}_{\rm m}$, we need to know the relation between $p$ and the quantities $h$ and $s$ and obtain their variations with respect to the metric. To do so, following the standard method, we introduce the Taub vector as $v^{\mu}= h u^{\mu}$. It should be noted that the Taub vector is defined by five scalar velocity-potential fields ($\phi, \alpha, \beta, \theta, s$) that are independent of the metric tensor (see Ref.~\cite{Schutz1970} for the physical meaning of these potentials). Namely, in the velocity-potential representation, the Taub vector is expressed as 
\begin{equation} \label{def:taub}
v_{\mu}=\partial_{\mu} \phi+\alpha \, \partial_{\mu} \beta+\theta \, \partial_{\mu} s. 
\end{equation}
According to this definition, one can obtain  \begin{equation} \label{def:h2}
 h^2=- g^{\mu\nu} v_{\mu} v_{\nu},
\end{equation}
and take the variation of the above relation with respect to the inverse metric $g^{\mu\nu}$, and reach the following relation
\begin{equation}\label{eq1}
2 h \delta h= - v_{\mu} v_{\nu}  \delta g^{\mu\nu},
\end{equation}
where we have used the fact that $v_{\mu}$ does not depend on the metric tensor. Substituting the definition of $v_{\mu}$ back into the above expression yields
\begin{equation}  \label{var-enth}
  \frac{\delta h}{\delta g^{\mu\nu}}= - \frac{h}{2} u_{\mu} u_{\nu}.  
\end{equation}
Furthermore, the first law of thermodynamics written as ${\rm d} p=n \,{\rm d} h-n\,\mathcal{T}\,{\rm d} s$~\cite{Rezzolla:2013}, reveals that $\frac{\partial p}{\partial h}\big|_s=n$ with $\mathcal{T}$ being the temperature. Hence, applying the constraint $\delta s=0$, we obtain $\delta p=-\frac{1}{2} n h u_{\mu} u_{\nu}\delta g^{\mu\nu}$, which corresponds to the pressure variation given in Eq.~\eqref{varpGR}. Accordingly, the variation of the action of the perfect fluid reduces to
\begin{equation}
\begin{aligned}
\delta  \mathcal{S}_{\rm m}= \int {\rm d^4}x \bigg[-\frac{1}{2} p \sqrt{-g} g_{\mu\nu} \delta g^{\mu\nu}-\frac{1}{2} \sqrt{-g} n h u_{\mu} u_{\nu} \delta g^{\mu \nu}\bigg],
\end{aligned}
\end{equation}
which can be rewritten as $\delta  \mathcal{S}_{\rm m}=-   \frac{1}{2} \sqrt{-g} \int {\rm d^4}x \, T_{\mu\nu}^{\rm pf}  \delta  g^{\mu \nu}$, where $T_{\mu\nu}^{\rm pf}= n h u_{\mu} u_{\nu} + p g_{\mu \nu}$. This result then matches the EMT of the perfect fluid given in Eq.~\eqref{em}.

We have shown that the variation of the action $\mathcal{S}_{\rm m}=\int {\rm d}^4x\, \sqrt{-g}\, p $ with respect to $g^{\mu\nu}$ provides us with the EMT of the perfect fluid. However, this action can also be varied with respect to other dynamical variables in order to obtain the full equations of motion for the perfect fluid and these equations imply that the divergence of the perfect fluid EMT vanishes. To begin with, we consider the variation of the action with respect to $\phi$ as follows;
\begin{align}
\delta \mathcal{S}_{\rm m}=\int {\rm d}^4x\, \frac{\delta{(\sqrt{-g}\, p)} }{\delta \phi} \delta \phi,   
\end{align}
which can be written as
\begin{align}
\delta \mathcal{S}_{\rm m}=\int {\rm d}^4x\, \left\{\frac{\partial{(\sqrt{-g}\, p)} }{\partial \phi}+\partial_{\mu}\left[\frac{\partial{(\sqrt{-g}\, p)} }{\partial (\partial_{\mu}\phi)}\right] \right\} \delta \phi.  
\end{align} 
We recall that $\frac{\partial p}{\partial h}\big|_s=n$, and according to Eqs.~\eqref{def:taub} and~\eqref{def:h2}, the first term in the above expression is zero since the Taub vector depends on the derivative of $\phi$ but not $\phi$ itself. Thus, using these, we obtain
\begin{align}
\delta \mathcal{S}_{\rm m}=\int {\rm d}^4x\, \partial_{\mu}\left[\sqrt{-g} n \frac{\partial{h} }{\partial (\partial_{\mu}\phi)}\right]  \delta \phi,  
\end{align} 
and Eq.~\eqref{def:h2} eventually implies\footnote{The covariant derivative of a vector can be written in terms of the partial derivative as $\nabla_{\mu} V^{\mu}=\frac{1}{\sqrt{-g}} \partial_{\mu}(\sqrt{-g} V^{\mu})$, see Eq. (3.34) in Ref.~\cite{Carroll:2004st}}
\begin{equation}  \label{eom1}
    \nabla_{\mu}(nu^{\mu})=0.
\end{equation}
On the other hand, varying the same action with respect to $\theta$ yields
 \begin{align}
\delta \mathcal{S}_{\rm m}=\int {\rm d}^4x\, \frac{\partial{(\sqrt{-g}\, p)} }{\partial \theta} \delta \theta=\int {\rm d}^4x\, \sqrt{-g} n \frac{\partial{h} }{\partial \theta}  \delta \theta, \end{align}
since $v_{\mu}$ does not depend on the derivative of $\theta$ and using Eq.~\eqref{def:h2} gives rise to
\begin{equation}    \label{eom2}
u^{\mu} \partial_{\mu}s=0.
\end{equation}
 The variations with respect to $\alpha, \beta$, and $s$ can be calculated in the same manner and related equations of motion correspondingly read
 \begin{align}
u^{\mu} \partial_{\mu}\beta=0\quad,\quad 
u^{\mu} \partial_{\mu}\alpha=0\quad\textnormal{and}\quad
u^{\mu} \partial_{\mu}\theta=\mathcal{T}  \label{eom5}.
\end{align}

As the final point, we would like to briefly discuss how the conservation of the perfect fluid EMT is deduced from the variational principle via the diffeomorphism invariance of the action of GR~\cite{Carroll:2004st,Felice:1992,Wald:1984rg}. Thanks to that the gravitational and matter parts of the action can be isolated in GR, the gravitational part being invariant under diffeomorphisms implies that so is the matter part. The variation of the matter part of the action $\mathcal{S}_{\rm m}[g^{\mu\nu}, \Psi^{i}]$ under a diffeomorphism can be written as
\begin{equation}  \label{varSm}
\delta \mathcal{S}_{\rm m}= \int {\rm d}^4x\, \frac{\delta(\sqrt{-g} \mathcal{L}_{\rm m})}{\delta g^{\mu\nu}}\delta g^{\mu\nu}+\int {\rm d}^4x\,\frac{\delta (\sqrt{-g} \mathcal{L}_{\rm m})}{\delta \Psi^{i}} \delta \Psi^{i},
\end{equation}
where $\Psi^{i}$ is a set of matter fields. As the matter action defined in Eq.~\eqref{matteraction} satisfies the equations of motion given in Eqs.~\eqref{eom1},~\eqref{eom2}, and~\eqref{eom5} for $\mathcal{L}_{\rm m}=p$, the second term vanishes through $\frac{\delta (\sqrt{-g} \mathcal{L}_{\rm m})}{\delta \Psi^{i}}=0$. Then, this condition leads to the fact that the first term also vanishes. The infinitesimal change in the metric is obtained from its Lie derivative along $\xi^{\mu}$ as $\mathcal{L}_{\xi} g^{\mu\nu}=\nabla^{\mu} \xi^{\nu}+\nabla^{\nu} \xi^{\mu}$, where $\xi^{\mu}$ is the infinitesimal vector field  generating the diffeomorphism. Substituting this relation in Eq.~\eqref{varSm}, we have 
\begin{equation}
 \int {\rm d}^4x\, \frac{\delta (\sqrt{-g} \mathcal{L}_{\rm m})}{\delta g^{\mu\nu}} 2 \nabla^{\mu} \xi^{\nu}=0,
 \end{equation}
and then making use of the definition given in Eq.~\eqref{gen-tmunudef} and applying Gauss' theorem~\cite{Dirac}, we finally obtain
\begin{equation}
\int {\rm d}^4x\, \sqrt{-g}\, \xi^{\nu} \nabla^{\mu} T_{\mu\nu}^{\rm pf}=0.  
\end{equation}
If this expression is demanded to hold for any arbitrary vector field $\xi^{\mu}$, it then implies $\nabla^{\mu} T_{\mu\nu}^{\rm pf}=0$, which is the local conservation of the perfect fluid EMT when Eqs.~\eqref{eom1},~\eqref{eom2}, and~\eqref{eom5} are the equations of motion under consideration. Therefore, it is evident from Eq.~\eqref{varSm} that defining the perfect fluid EMT through Eq.~\eqref{tmunudef} does not require by alone its conservation, but the validity of the aforementioned equations of motion must accompany it. In other words, in matter-type modifications, the equations of motion given in Eqs.~\eqref{eom1},~\eqref{eom2}, and~\eqref{eom5} also change with the addition of extra terms coming from $f$ part of the total matter Lagrangian density, and hence, the intact part of the EMT, $T_{\mu\nu}^{\rm pf}$ is not necessarily conserved.

\subsubsection{The matter Lagrangian density $\mathcal{L}_{\rm m}=- \rho$}  \label{sec:Lm-rho}
In this case, we let the EoS of the perfect fluid be given as $\rho=\rho(n,s)$ and the matter Lagrangian density be defined as $\mathcal{L}_{\rm m}=- \rho$.  So, the action of the perfect fluid reads $\mathcal{S}_{\rm m}=\int {\rm d}^4x\, \sqrt{-g}\, (-\rho) $.
In a similar fashion to the previous section, to obtain $\delta \rho=\frac{\partial \rho}{\partial n}\big|_s \delta n +\frac{\partial \rho}{\partial s}\big|_n \delta s $, we need to know the variations of $n$ and $s$, as well as the relation between $\rho$ and these quantities. We define the particle number flux vector density $n^{\mu}$ as $n^{\mu}= \sqrt{-g} n u^{\mu}$ implying
\begin{equation}\label{n2}
  n^2=\frac{g_{\mu\nu} n^{\mu}n^{\nu}}{g}. 
\end{equation}
Regarding the variations of $n$ and $s$, we consider two necessary constraints: (i) $\delta n^{\mu}=0$ and (ii) $\delta s=0$, see Ref.~\cite{Felice:1992} for details. 
To obtain the relation between $\rho$ and two independent quantities $n$ and $s$, one can utilize the first law of thermodynamics. 

Now we employ the variational procedure. Varying Eq.~\eqref{n2} with respect to the inverse metric $g^{\mu\nu}$ and applying the assumption $\delta n^{\mu}=0$, we obtain
\begin{equation}
2 n \delta n= -n_{\mu} n_{\nu} \left(\frac{1}{g} \delta g^{\mu\nu}+\frac{1}{g^2} g^{\mu\nu} \delta g \right).   
\end{equation}
Next, making use of $\delta g=- g g_{\sigma\epsilon} \delta g^{\sigma\epsilon}$ and substituting the definition of $n_{\mu}$ into the above expression give
\begin{equation}  \label{var-pn}
  \frac{\delta n}{\delta g^{\mu\nu}}= \frac{n}{2} (u_{\mu} u_{\nu}+g_{\mu\nu}).  
\end{equation}
Using an alternative expression of the first law of thermodynamics given as ${\rm d} \rho=h \,{\rm d} n+n\,\mathcal{T}\,{\rm d} s$, we have $\frac{\partial \rho}{\partial n}\big|_s=h$ and $\frac{\partial \rho}{\partial s}\big|_n=n\mathcal{T}$. Therefore, after applying the constraints, we obtain $\delta \rho= \frac{1}{2}n\,h (u_{\mu} u_{\nu}+g_{\mu\nu}) \delta g^{\mu\nu}$ which corresponds to the energy density variation given in Eq.~\eqref{varrhoGR}. Using this along with  $\delta \sqrt{-g}=-\frac{1}{2} \sqrt{-g} g_{\mu\nu} \delta g^{\mu\nu}$, finally we reach
\begin{equation}
\begin{aligned}
  \delta  \mathcal{S}_{\rm m}= \int {\rm d^4}x \bigg[&\frac{1}{2} \rho \sqrt{-g} g_{\mu\nu} \delta g^{\mu\nu}  \\
  &-\frac{1}{2} \sqrt{-g} n h    (u_{\mu} u_{\nu}+g_{\mu\nu}) \delta g^{\mu \nu}\bigg].
\end{aligned}
\end{equation}
This expression can also be written as $\delta  \mathcal{S}_{\rm m}=-  \frac{1}{2} \sqrt{-g} \int {\rm d^4}x \, T_{\mu\nu}^{\rm pf}  \delta  g^{\mu \nu}$, where $T_{\mu\nu}^{\rm pf}= n h u_{\mu} u_{\nu} + (n h-\rho) g_{\mu \nu}$ is equivalent to the EMT of the perfect fluid given in Eq.~\eqref{em}. We should also note that the above discussion regarding the conservation of the perfect fluid EMT is also valid for $\mathcal{L}_{\rm m}=-\rho$ case, which satisfies the same equations of motion~\eqref{eom1},~\eqref{eom2}, and~\eqref{eom5}.

\subsection{The second metric derivative of the matter Lagrangian density for perfect fluid}
\label{sec:secderlm}
 As mentioned before, in perfect fluid case, Eqs.~\eqref{varpGR} and~\eqref{varrhoGR} reveal that taking their second metric derivative brings the variations of energy density and pressure together in the same expression. This is a situation that has not been encountered until the attempts to modify the matter Lagrangian density of EH action. One immediate solution that comes to mind may be to use the variations given in Eqs.~\eqref{varpGR} and~\eqref{varrhoGR} simultaneously. In general, this is not physically acceptable since these two variations are derived from different thermodynamic assumptions\footnote{In Ref.~\cite{Haghani:2023uad}, these two variations are used at the same time to find the expression $\frac{\partial^2 \mathcal{L}_{\rm m}}{\partial g^{\mu \nu} \partial g^{\sigma\epsilon}}$. However, this can only be done through the new interpretation presented in the current paper, i.e., as a freedom in determining the form of interaction in nonminimally interacting models. Having said that, without the aforementioned interpretation there is a physical ambiguity in their simultaneous use. To illustrate this, we add Eqs.~\eqref{varpGR} and~\eqref{varrhoGR}, and obtain
\begin{align*}
 \frac{\delta (\rho+p)}{\delta g^{\mu\nu}}=\frac{1}{2}(\rho+p)g_{\mu\nu}.
\end{align*}
Multiplying both sides by $-\sqrt{-g}$, and then using the relation $\delta \sqrt{-g}=-\frac{1}{2} \sqrt{-g} g_{\mu\nu} \delta g^{\mu\nu}$ on the right-hand side lead to
\begin{align*}
\delta \ln{(\rho+p)}=-\delta \ln{\sqrt{-g}} \quad \longrightarrow \quad (\rho+p)\sqrt{-g}=\rm const.
\end{align*} 
This means that their simultaneous use corresponds to a fixed EoS. Consequently, using these variations simultaneously in the second derivative term is no more advantageous than removing this term in point of that the Lagrangian formulation must still be compromised.} as discussed in Sections~\ref{sec:Lm-rho} and~\ref{sec:Lm-p}. In other words, in the case of $\mathcal{L}_{\rm m}=p(h,s)$, two independent variables are $h$ and $s$, and hence $p$ and $\rho$ are not independent and determined from these two variables ($h,s$).
Indeed at constant entropy, the energy density is defined by 
\begin{equation}\label{firstlaw1}
\rho=h \frac{\partial p}{\partial h}\Big|_s-p. 
\end{equation}
Similarly, in the case of $\mathcal{L}_{\rm m}=-\rho(n,s)$, two independent variables are $n$ and $s$, and thereby, $\rho$ and $p$ are not independent and determined from the two variables ($n,s$). At constant entropy, the pressure is defined as 
\begin{equation}\label{firstlaw2}
p=n \frac{\partial \rho}{\partial n}\Big|_s-\rho.
\end{equation}

In light of the above discussion, we can deduce that in $\mathcal{L}_{\rm m}=p$ case, $\rho$ is defined by $p$, i.e.,  $\rho=\rho(p)$ and in $\mathcal{L}_{\rm m}=-\rho$ case, $p$ is defined by $\rho$, i.e., $p=p(\rho)$ as we usually consider barotropic fluids in the context of cosmology and astrophysics. Therefore, the proper way of calculating the second metric derivative of the matter Lagrangian density is to make use of the EoS of the corresponding fluid. Now, we will present this calculation for both cases and discuss the possible consequences of it. 

\subsubsection{The choice of $\mathcal{L}_{\rm m}=p$}
Although the second metric derivative of pressure can be calculated directly from Eq.~\eqref{varpGR}, to shed light on the assumptions behind the calculation, let us take a step back and begin with the results presented in Sec.~\ref{sec:Lm-p}.  In that section, we have obtained that $\frac{\delta p(h, s)}{\delta g^{\mu\nu}}=-\frac{1}{2} h\,\frac{\partial p}{\partial h}\big|_s u_{\mu} u_{\nu}$. We should mention that to achieve this result, the constraint $\delta s=0$ has been applied. Keeping these in mind, the second metric derivative of the pressure can be written as
\begin{align}\label{par2P}
\nonumber
 \frac{\partial^2 p}{\partial g^{\mu \nu} \partial g^{\sigma\epsilon}} =-\frac{1}{2}\bigg[&\bigg(\frac{\partial h}{\partial g^{\mu\nu}}\frac{\partial p}{\partial h}+h\frac{\partial}{\partial g^{\mu\nu}}\frac{\partial p}{\partial h}\bigg)u_{\sigma}u_{\epsilon}\\
 &+h\,\frac{\partial p}{\partial h}\frac{\partial (u_{\sigma}u_{\epsilon})}{\partial g^{\mu\nu}}\bigg].
\end{align}  
Using Eq.~\eqref{var-enth}, as well as the following relation
\begin{align}\label{paru2}
\frac{\partial (u_{\sigma}u_{\epsilon})}{\partial g^{\mu\nu}}=u_{\sigma}u_{\epsilon} u_{\mu}u_{\nu},
\end{align}
 obtained from the normalization condition $g^{\sigma\epsilon}u_{\sigma} u_{\epsilon}=-1$, the first and third terms of Eq.~\eqref{par2P} can be simplified without any additional assumptions. Now, we need to obtain the second term, and to do so, it is necessary to reconsider the previously applied constraint, viz., the conservation of the specific entropy during the variation, $\delta s=0$. Regarding this, one can then deduce that
 \begin{align}\label{parp2}
\frac{\partial}{\partial g^{\mu\nu}}\frac{\partial p}{\partial h}=\frac{\partial^2 p}{\partial h^2}\Big|_s\frac{\partial h}{\partial g^{\mu\nu}}.
\end{align}
Hence, the next task is to find the term $\frac{\partial^2 p}{\partial h^2}\big|_s$. From Eq.~\eqref{firstlaw1}, it is straightforward to calculate that
\begin{align}\label{par2p=par1rho}
\frac{\partial^2 p}{\partial h^2}\Big|_s=\frac{1}{h}\frac{\partial \rho}{\partial h}\Big|_s.
\end{align}
Inserting Eqs.~\eqref{var-enth},~\eqref{paru2}, and~\eqref{par2p=par1rho} into Eq.~\eqref{par2P}, and after some simplification, we obtain 
\begin{align}\label{secderofp1}
 \frac{\partial^2 p}{\partial g^{\mu \nu} \partial g^{\sigma\epsilon}} = - \frac{h}{4} \left(\frac{\partial p}{\partial h}\Big|_s-\frac{\partial \rho}{\partial h}\Big|_s \right)u_{\mu} u_{\nu} u_{\sigma} u_{\epsilon}.
 \end{align}
Eventually, considering the EoS $\rho=\rho(p)$ which implies $\frac{\partial \rho}{\partial h}|_s=\frac{\partial \rho}{\partial p}|_s \frac{\partial p}{\partial h}|_s$, one can find the second metric variation of $\mathcal{L}_{\rm m}=p$ as
\begin{align} \label{del2p}  
 \frac{\partial^2 \mathcal{L}_{\rm m}(=p)}{\partial g^{\mu \nu} \partial g^{\sigma\epsilon}} = - \frac{1}{4} \left(1-\frac{\partial \rho}{\partial p}\Big|_s\right)(\rho+p) u_{\mu} u_{\nu}u_{\sigma} u_{\epsilon},
 \end{align}
in which Eq.~\eqref{firstlaw1} is also substituted. 

As seen, in the final result, the first derivative of $\rho$ with respect to $p$ appears, which is indeed related to the definition of the sound speed in the fluid, i.e., $\frac{\partial p}{\partial \rho} \big|_{s}=c_{\rm s}^2$. It is obvious that Eq.~\eqref{del2p} diverges when we consider dust ($p=0$ and $c_{\rm s}^2=0$). Based on this point, it is not possible to derive the field equations that are valid in the presence of dust from $\mathcal{L}_{\rm m}^{\rm tot}=p+f$ when $f$ is an arbitrary function of $g_{\mu\nu}T^{\mu\nu}$ or $T_{\mu\nu}T^{\mu\nu}$. Therefore, it turns out to be reasonable to omit this second derivative term for the case $\mathcal{L}_{\rm m}=p$. Furthermore, one may interpret the omission of this term as if it is taken to be zero. 
Regarding the above expression given in Eq.~\eqref{del2p}, obviously, it vanishes when $\frac{\partial \rho}{\partial p}=1$. However, it is worth noting that this result is not carried to other $\rho$ and $p$ terms in the full expression of $\theta_{\mu\nu}$ and only valid in the expression~\eqref{del2p} [see Eqs.~\eqref{theta-EMSF} and~\eqref{theta-pf}], thereby, it does not mean that we fix the EoS of the usual fluid. For this reason, it is more appropriate to regard this choice as omitting the second metric derivative of $\mathcal{L}_{\rm m}$ rather than as setting it to zero. We take advantage of this choice to present a well-defined interaction model that can properly describe the beloved fluid, dust, in cosmology. However, as mentioned earlier, this choice comes at the price of compromising the Lagrangian formulation of the interaction model. We should note that this issue arises not only in GR with nonminimal interactions constructed from the scalars $g_{\mu\nu} T^{\mu\nu}$ and $T_{\mu\nu} T^{\mu\nu}$ but also in extended $f(R)$, teleparallel gravity theories by incorporating the same scalars and in theories with matter-curvature couplings like $f(R_{\mu\nu} T^{\mu\nu})$~\cite{Haghani:2013oma,Odintsov:2013iba}, $f(G_{\mu\nu} T^{\mu\nu})$~\cite{Asimakis:2022jel}. Since these theories also assume the term $\frac{\partial^2 \mathcal{L}_{\rm m}}{\partial g^{\mu \nu} \partial g^{\sigma\epsilon}}$ emerging from the variation of their actions as zero in the presence of perfect fluid, they work properly only at the level of field equations. We would like to add that another choice might be making use of the variations in Eqs.~\eqref{varpGR} and~\eqref{varrhoGR} simultaneously as is studied in Ref.~\cite{Haghani:2023uad} but the derivation of the field equations from $\mathcal{L}_{\rm m}^{\rm tot}=p+f$ is still not valid. In this case, using the variations simultaneously does not have any physical implications other than determining the form of the interaction, as in our choice to remove the second metric derivative of $\mathcal{L}_{\rm m}$.

\subsubsection{The choice of $\mathcal{L}_{\rm m}=- \rho$}
In the previous section, it is demonstrated that choosing $\mathcal{L}_{\rm m}=p$, there is a divergence in the inclusion of dust in these models at the level of Lagrangian descriptions, which is firmly resolved at the level of field equations. On the other hand, there are other choices of $\mathcal{L}_{\rm m}$ to define a perfect fluid. Although they properly describe the perfect fluid, as discussed before, the other cases are less commonly used in the literature compared to the well-known case $\mathcal{L}_{\rm m}=p$. Now, we focus our attention on the case $\mathcal{L}_{\rm m}=-\rho$ to investigate whether there is a well-defined Lagrangian formulation for the interaction models under consideration. Nevertheless, we should emphasize that even with a well-defined matter Lagrangian density, these models are not new theories, but other forms of nonminimal interactions in GR. 

In a similar manner to the previous calculation, we begin with the result given in Sec.~\ref{sec:Lm-rho}.  The second metric derivative of energy density is calculated from $\frac{\delta \rho(n,s)}{\delta g^{\mu\nu}}= \frac{1}{2}n\,\frac{\partial \rho}{\partial n}\big|_s (u_{\mu} u_{\nu}+g_{\mu\nu})$ as follows:
\begin{equation} 
\begin{aligned} 
\label{secvar2}
 \frac{\partial^2 \rho}{\partial g^{\mu \nu} \partial g^{\sigma\epsilon}}=  \frac{n}{4}  \bigg[ &\left(\frac{\partial \rho}{\partial n}\Big|_s+n\frac{\partial^2 \rho}{\partial n^2}\Big|_s\right)  \\ &\times (u_{\mu} u_{\nu}+g_{\mu\nu}) (u_{\sigma} u_{\epsilon}+g_{\sigma\epsilon})\\
 &+2 \frac{\partial \rho}{\partial n}\Big|_s \big(u_{\sigma} u_{\epsilon}u_{\mu} u_{\nu}-g_{\sigma\mu}g_{\epsilon\nu}\big)\bigg].
\end{aligned}
\end{equation}
To simplify the above relation, we utilize Eqs.~\eqref{var-pn} and~\eqref{paru2} and the relation $\frac{\partial g_{\sigma\epsilon}}{\partial g^{\mu\nu}}=-g_{\sigma\mu}g_{\epsilon\nu}$. The second derivative term in Eq.~\eqref{secvar2} can be calculated easily after using the expression~\eqref{firstlaw2}, so as to obtain 
\begin{align}\label{del2rho}
n\frac{\partial^2 \rho}{\partial n^2}\Big|_s=\frac{\partial p}{\partial n}\Big|_s.
\end{align}
 Finally, considering  the EoS $p=p(\rho)$ which implies $\frac{\partial p}{\partial n}|_s=\frac{\partial p}{\partial \rho}|_s \frac{\partial \rho}{\partial n}|_s$ and then substituting Eqs.~\eqref{firstlaw2} and~\eqref{del2rho}, we obtain the second metric derivative of $\mathcal{L}_{\rm m}=- \rho$ as
\begin{equation}
\begin{aligned} 
\label{del2rhowg}
 \frac{\partial^2 \mathcal{L}_{\rm m}(=- \rho)}{\partial g^{\mu \nu} \partial g^{\sigma\epsilon}} =- \frac{1}{4} (\rho+p) \bigg[&
\left(1+\frac{\partial p}{\partial \rho}\Big|_s\right)  \\
&\times (u_{\mu} u_{\nu}+g_{\mu\nu}) (u_{\sigma} u_{\epsilon}+g_{\sigma\epsilon})\\
&+2 (u_{\sigma} u_{\epsilon} u_{\mu} u_{\nu}-g_{\sigma \mu} g_{\epsilon\nu}) \bigg].
 \end{aligned}
\end{equation}
 Unlike the previous case, it can be seen that there is no ill-defined term even in the presence of dust for the case $\mathcal{L}_{\rm m}=- \rho$. Therefore, we continue without removing the second derivative of $\mathcal{L}_{\rm m}=- \rho$. By doing so, we find
\begin{equation}\label{thetarho}
 \theta_{\mu\nu}^{\rm pf}=p\,(\rho+p) \left[1+3 \frac{\partial p}{\partial \rho}\right] (u_{\mu} u_{\nu}+g_{\mu\nu}), 
\end{equation}  
after substituting Eqs.~\eqref{em} and~\eqref{del2rhowg} into the definition~\eqref{theta}.

Recall that in Sec.~\ref{sec:nonmin}, we discussed how matter-type theories are equivalent to nonminimal interaction models in GR. Focusing on the specific theory of EMSG, we derived the definition of the EMSF from this new perspective, under the assumption that the matter Lagrangian density of the usual field is independent of the derivatives of the metric tensor. In the following sections, we will calculate the EMSF and the corresponding field equations of the model for when the usual field is a perfect fluid. We will then revisit the cosmological models of EMSG to provide insights into how our new findings can be applied and interpreted in various contexts of EMSG and similar theories.

\section{Reconsidering EMSG cosmology in the context of EMSF approach} \label{emsg-cosm}

We proceed by considering perfect fluid as the usual material field in EMSF interaction. In line with existing literature, we choose $\mathcal{L}_{\rm m}=p$, and by substituting Eq.~\eqref{em} into Eq.~\eqref{theta-EMSF}, we derive the following expression:
\begin{equation} \label{theta-pf}
\theta_{\mu\nu}^{\rm emsf,pf}=-(\rho+p) (\rho+3p) u_{\mu} u_{\nu}, 
\end{equation}
the usual expression used for the perfect fluid in the EMSG models studied so far. Using Eq.~\eqref{theta-pf} in Eq.~\eqref{eq:int1}, the corresponding interaction kernel for the perfect fluid reads 
\begin{equation}
\begin{aligned}
\label{eq:Q}
  Q_{\nu}=&\nabla_{\nu}f+2 \,(\rho+p) (\rho+3p) u_{\mu} u_{\nu} \nabla^{\mu}f_{\mathbf{T^2}} \\
  &+2 f_{\mathbf{T^2}} \nabla^{\mu}\left[(\rho+p) (\rho+3p) u_{\mu} u_{\nu}\right].
  \end{aligned}
\end{equation}
From the definition given in Eq.~\eqref{def-EMSF} along with Eq.~\eqref{theta-pf}, in the presence of perfect fluid as the usual material field, the EMT of the accompanying EMSF has the following form
\begin{equation}
 \begin{aligned}
  \label{cmunudef}
 T_{\mu\nu}^{\rm emsf}= f g_{\mu\nu}+2 f_{\mathbf{T^2}} (\rho+p) (\rho+3p) u_{\mu} u_{\nu}.
 \end{aligned}
 \end{equation}
Also, by combining the contribution of the usual material field with that of EMSF, from Eq.~\eqref{emt-tot}, we obtain
 \begin{align}
  &\rho_{\rm tot}= \rho-f+2 f_{\mathbf{T^2}} (\rho+p) (\rho+3p), \\
  &p_{\rm tot}=p+f.
 \end{align}
Having established the proper definition of the EMSF in alignment with the field equations found in existing literature on EMSG, we will, in the next two sections, delve into specific applications of the EMSF interaction model within the realm of cosmology.

\subsection{EMSF in the Dark Sector}\label{EMSF_cos}
In the framework of EMSF interaction, one can construct models in which components, whose nature is not well understood yet, like CDM and relativistic relics couple to the spacetime under the influence of their interactions with EMSF whereas well-known sources such as baryons and photons couple to the curvature with their usual EMTs only. Although it seems that these species interact among themselves through EMSF, from one perspective, we can interpret the contributions coming from this field/interaction as a dark energy (DE) component. Depending on the form of the interaction determined by $f$, these DE models have effects on early and late times of the Universe, hence may ameliorate the current tensions such as $H_0$ tension~\cite{Verde:2019ivm,Riess:2021jrx,Riess:2019qba,DiValentino:2020zio,DiValentino:2021izs,Kamionkowski:2022pkx,DiValentino:2022fjm} and $S_8$ tension~\cite{DiValentino:2020vvd,Douspis:2018xlj,Abdalla:2022yfr} within the $\Lambda$CDM model. In what follows, we will investigate this aspect of the EMSF interaction by making use of different forms of the $f$ function.

\subsubsection{Interacting dark energy (IDE) models}
It is possible with EMSF to construct IDE models~\cite{Bolotin:2013jpa,Wang:2016lxa} comprising DE nonminimally interacting with dark matter (DM). These models have recently gained an increased interest in addressing some cosmological tensions, and moreover, in Ref.~\cite{Escamilla:2023shf}, model-independent reconstruction of the IDE kernel has been performed. In common phenomenological models, $\nu=0$ component of the interaction kernel ($Q_0$) is in general assumed to be a simple function of energy density $\rho$ and of Hubble radius $H^{-1}$, and the corresponding Taylor expansion at the first order can be written as $Q_0=3H(\zeta_1 \rho_{\rm c}+\zeta_2 \rho_{\rm de})$ where $\rho_{\rm c}$ and $\rho_{\rm de}$ are energy densities of CDM and DE respectively with $\zeta_1$ and $\zeta_2$ being free parameters. On top of that, it is later realized that these phenomenological models primarily alleviate the $H_0$ tension~\cite{Bernui:2023byc,Zhai:2023yny} (as well as solving cosmic coincidence problem on the theoretical side~\cite{Bolotin:2013jpa,Wang:2016lxa}), while $S_8$ tension is usually exacerbated. Also, they suffer from perturbation instability~\cite{zhao,Majerotto,He} and early time instabilities that can be avoided by setting the EoS parameter of DE to $w_{\rm de} \equiv \frac{p_{\rm de}}{\rho_{\rm de}}=-0.9999$ with $p_{\rm de}$ being pressure of it. 

On the other hand, Eq.~\eqref{eq:int1} shows that EMSF in the dark sector may generate IDE models having extremely intricate interaction forms even with the most straightforward functions chosen for $f$. The analysis in Ref.~\cite{Escamilla:2023shf} indicates slightly oscillatory dynamics in the interaction kernel, thereby, a sign change in the direction of the energy transfer between DE and DM, and a possible transition from $\rho_{\rm de}<0$ in early times to $\rho_{\rm de}>0$ at late times of the Universe. Since such a nontrivial dynamics is difficult to achieve via simple interaction kernels, it is apparent that the EMSF in the dark sector in this sense deserves attention.

We consider the FLRW spacetime metric with flat spacelike sections, ${\rm d}s^{2}=-{\rm d}t^{2}+a^{2}{\rm d}{\vec r}^{2}$, where the scale factor $a = a(t)$ is a function of cosmic time $t$ only. Assuming constant EoS parameters for barotropic perfect fluids, viz., $\frac{p}{\rho}\equiv w=\rm const.$, we consider that CDM (c) interacts nonminimally with EMSF, whereas known sources, viz., baryon (b) and radiation (r) have no EMSF partners, then the field equations read
\begin{align}
3 H^2=&\kappa \left(\rho_{\rm r}+\rho_{\rm b}+\rho_{\rm c} -  f + \rho_{\rm c} \frac{{\rm d} f}{{\rm d} \rho_{\rm c}}\right),  \\
-2\dot{H}-3H^2=&  \kappa \left(\frac{1}{3}  \rho_{\rm r}+ f \right),
\end{align}
where $H=\frac{\dot{a}}{a}$ is the Hubble parameter and the overdot denotes the
derivative with respect to $t$. In these relations, we can interpret the extra terms coming from the EMSF as effective DE with the following energy density and pressure 
\begin{align}
\rho_{\rm de}=-f+ \rho_{\rm c} \frac{{\rm d} f}{{\rm d} \rho_{\rm c}} \quad , \quad p_{\rm de}=f.
\end{align}
Hence, the EoS parameter of DE reads
\begin{align}
 w_{\rm de}=\left(-1+\frac{{\rm d} \ln{f}}{{\rm d} \ln{\rho_{\rm c}}}\right)^{-1},   
\end{align}
 which implies that the model can generate a dynamical DE, i.e., a DE that has an EoS parameter evolving with cosmic time/redshift, depending on the choice for the function $f$. However, if $f(\rho_{\rm c})$ is flat enough, viz., $\frac{{\rm d}f}{{\rm d}\rho_{\rm c}}\sim 0$, then $w_{\rm de}\sim-1$. The part of the EMT describing the dark sector is conserved within itself, and the continuity equation reads as follows;
\begin{equation} \label{cont-emsf}
\dot{\rho}_{\rm c}\left(1+ \rho_{\rm c} \frac{{\rm d}^2 f}{{\rm d} \rho_{\rm c}^2}\right)+3 H \rho_{\rm c} \left(1+\frac{{\rm d} f}{{\rm d} \rho_{\rm c}}\right)=0,
\end{equation}
also, setting $\nu=0$ in Eq.~\eqref{eq:Q}, the interaction kernel $Q_0$ turns out to be\footnote{In conventional models, the interaction terms are generally of the forms $Q_0\propto \gamma H\rho_1+\Gamma H\rho_2$, $Q_0\propto \gamma H\rho_1$, and $Q_0\propto \Gamma H\rho_2$, with $\gamma$ and $\Gamma$ being arbitrary constants while the term $\dot{\rho}$ is not commonly used. In fact, it is seen with a straightforward rescaling that $\dot{\rho}$ gives rise to nothing but interaction kernels more generic than the linear function of $\rho$~\cite{ABK}.}
\begin{equation}
  Q_0=  \rho_{\rm c} \left(\dot{\rho}_{\rm c}  \frac{{\rm d}^2 f}{{\rm d} \rho_{\rm c}^2}+3 H \frac{{\rm d} f}{{\rm d} \rho_{\rm c}}\right).
\end{equation}
To demonstrate these points, we proceed with EMPF~\cite{Akarsu:2017ohj,Board:2017ign}, the powered form of EMSF, described by 
\begin{equation}
f=\alpha \rho_{\rm c}^{2\eta}
\end{equation}
in the presence of CDM. So, from Eq.~\eqref{cont-emsf}, we obtain
\begin{equation}
\dot{\rho}_{\rm c}\left[1+ \alpha 2\eta (2\eta-1)  \rho_{\rm c}^{2\eta-1}\right]+3 H \rho_{\rm c} \left(1+ \alpha 2\eta \rho_{\rm c}^{2\eta-1} \right)=0.    
\end{equation}
Thereby, the energy density and pressure of the DE are indeed arisen from the interaction between CDM and its accompanying EMPF as follows;
\begin{align}
\rho_{\rm de}= \alpha (2 \eta-1) \rho_{\rm c}^{2\eta}  \quad \textnormal{and} \quad p_{\rm de}=\alpha \rho_{\rm c}^{2\eta},
\end{align}
provided that $\eta\neq\frac{1}{2}$, accordingly giving 
\begin{align}
 w_{\rm de}=\frac{1}{2 \eta-1} =\rm const.,   
\end{align}
where $\alpha$ is a free parameter that determines the amount of EMPF with respect to CDM. Note that a value of $\eta=0$ corresponds to the $\Lambda$CDM model, while $\eta\sim0$ results in to a $w$CDM-like model. However, the underlying physics is entirely different in the sense that the accelerated expansion of the Universe is virtually due to the nonminimal interaction between the EMPF and CDM, rather than an isolated physical DE source for the cases with $\eta<1/2$, for which the EMPF contribution to the Friedmann equation is effective at low values of energy densities. Moreover, the energy densities of the CDM and of EMPF are conserved together, namely, CDM in general does not dilute as $\rho_{\rm c}\propto a^{-3}$ as the Universe expands, see Ref.~\cite{Akarsu:2017ohj} for details.

As can be seen, to achieve a dynamic DE model, it is necessary to extend beyond the power-law form of the EMSF interaction. For example, consider CDM interacting with EMLF described by the following expression: 
\begin{equation}
f=\alpha \ln{(\lambda \rho_{\rm c}^2})
\end{equation}
with $\alpha$ and $\lambda$ being constants, which gives rise to
\begin{align}
\rho_{\rm de}=-\alpha \ln{(\lambda \rho_{\rm c}^2)}+ 2 \alpha \quad , \quad p_{\rm de}=\alpha \ln{(\lambda \rho_{\rm c}^2)}.
\end{align}
If we choose $\lambda=\exp (-\Lambda/ \alpha) \rho_{\rm c0}^{-2}$, the model inherently incorporates a cosmological constant, $\Lambda$, in addition to its dynamical component. In this model, $\rho_{\rm c}$ has an altered scale factor dependency due to the nonconservation of the usual EMT interacting nonminimally with the EMLF, and hence, by crossing below zero at large redshifts, it can accommodate a mechanism for screening $\Lambda$ at this epoch, in line with suggestions for alleviating some of the cosmological discrepancies that arise within the standard $\Lambda$CDM model. See Ref.~\cite{Akarsu:2019ygx} for a detailed discussion of a slightly different cosmological model in which baryons are also included in the interaction and Ref.~\cite{Acquaviva:2022bju} for its dynamical analysis.
 
 Note that the resulting DE here has a constant inertial mass density, i.e., $\rho_{\rm de}+p_{\rm de}=2\alpha$, which has been studied in Ref.~\cite{Acquaviva:2021jov} under the name of \textit{simple-graduated DE} as an alternative to the usual cosmological constant having null inertial mass density, viz., $\rho_{\Lambda}+p_{\Lambda}=0$. Its observational analysis suggests that the inertial mass density of DE yields slightly positive values, viz., $\mathcal{O}(10^{-12})\,\rm eV^4$, though consistent with zero inertial mass density of the usual cosmological constant. This source  has recently been of interest to many as it can resemble $\Lambda$ today, while leading to a future singularity dubbed as the little sibling of the big rip (LSBR) for $\rho_{\rm de}+p_{\rm de}=\rm const < 0$ or a finite future bounce for $\rho_{\rm de}+p_{\rm de} = \rm const > 0$~\cite{Bouhmadi-Lopez:2014cca,Albarran:2016mdu,Bouali:2019whr}.

\subsubsection{EMSF acting as noncanonical scalar fields }

In line with Ref.~\cite{Akarsu:2018aro}, yet without choosing $f$ function, we assume that relativistic relics ($\nu$) described by $w=\frac{1}{3}$ interact nonminimally with EMSF, whereas usual sources, viz., dust (d) and photons ($\gamma$) couple to the curvature with their usual EMTs only. We also include a bare cosmological constant, $\Lambda$, in the model.  For the FLRW spacetime metric stated above, the field equations of this model read
 \begin{align}
3 H^2=&\Lambda+\kappa\left(\rho_{\rm d}+\rho_{\gamma}+\rho_{\nu} -  f +2  \rho_{\nu} \frac{{\rm d} f}{{\rm d} \rho_{\nu}}\right),  \\
-2\dot{H}-3H^2=&-\Lambda+\kappa \left(\frac{1}{3}  \rho_{\gamma}+ \frac{1}{3}  \rho_{\nu}+ f \right).
\end{align}
Following Ref.~\cite{Akarsu:2018aro}, we interpret along with $\Lambda$, the contribution coming from the interaction of relativistic species with their accompanying EMSF as an effective DE component whose energy density and pressure are
\begin{align}
\rho_{\rm de}=\rho_{\nu}-f+ 2 \rho_{\nu} \frac{{\rm d} f}{{\rm d} \rho_{\nu}} +\Lambda \quad , \quad p_{\rm de}=\frac{1}{3}  \rho_{\nu}+f-\Lambda,
\end{align}
and accordingly, the continuity equation for this sector reads
\begin{equation} \label{cont-nu}
\dot{\rho}_{\nu}\left(1+\frac{{\rm d} f}{{\rm d} \rho_{\nu}} +2 \rho_{\nu} \frac{{\rm d}^2 f}{{\rm d} \rho_{\nu}^2}\right)+2 H \rho_{\nu} \left(2+ 3 \frac{{\rm d} f}{{\rm d} \rho_{\nu}}\right)=0.    
\end{equation}
Hence, setting $\nu=0$ in Eq.~\eqref{eq:Q}, the corresponding interaction kernel is turned out be;
\begin{equation}
\label{cont-nuint}
  Q_0=  \dot{\rho}_{\nu} \frac{{\rm d} f}{{\rm d} \rho_{\nu}}+2 \rho_{\nu}\left( \dot{\rho}_{\nu} \frac{{\rm d}^2 f}{{\rm d} \rho_{\nu}^2}+3 H \frac{{\rm d} f}{{\rm d} \rho_{\nu}}\right).
\end{equation}

A nontrivial coupling of the scalar field to the standard model neutrinos has been proposed in Refs.~\cite{Sakstein:2019fmf,CarrilloGonzalez:2020oac} both to resolve the $H_0$ tension and to ameliorate the significant fine-tuning problems of the standard early dark energy (EDE) models, where a dark component effective around the epoch of matter-radiation equality is considered as a resolution to the $H_0$ tension~\cite{Poulin:2018cxd}. To apply this approach in the current interaction model, in Ref.~\cite{Akarsu:2018aro}, the authors consider the scale-independent EMSF with the choice of 
\begin{equation}
f=\frac{2}{\sqrt{3}}\alpha \rho_{\nu}
\end{equation}
in the presence of only relativistic species. The energy density and pressure of the DE composed of $\Lambda$ and relativistic species interacting with scale-independent EMSF become
\begin{align}
\rho_{\rm de}= \left(1+\frac{2\alpha}{\sqrt{3}}\right)\rho_{\nu}+\Lambda  \quad , \quad p_{\rm de}=\left(\frac{1}{3}+\frac{2\alpha}{\sqrt{3}}\right)\rho_{\nu}-\Lambda,
\end{align}
which leads to a dynamical EoS parameter as
\begin{equation}
\label{eq:eosnulambda}
w_{\rm de}=\frac{\left(\frac{1}{3}+\frac{2\alpha}{\sqrt{3}}\right)\rho_{\nu}-\Lambda}{\left(1+\frac{2\alpha}{\sqrt{3}}\right)\rho_{\nu}+\Lambda}.
\end{equation}
It is reminiscent of a canonical scalar field as $w_{\rm de}\sim \frac{\rho_{\nu}-\Lambda}{\rho_{\nu}+\Lambda}$ for $ \alpha \gg 0$ and, in general, of the one that has been obtained by introducing a noncanonical scalar field~\cite{Mukhanov:2005bu} particularly considered for unifying CDM and DE~\cite{Mishra:2018tki}. They have noted in Ref.~\cite{Akarsu:2018aro} that depending on whether the relativistic species or the cosmological constant are dominant, the EoS parameter given in Eq.~\eqref{eq:eosnulambda} ranges respectively between the following limits:
\begin{equation}
\begin{aligned}
w_{\rm{de},\nu}=\frac{1}{3}\left(1+\frac{4\alpha}{2\alpha+\sqrt{3}}\right) \quad \textnormal{and}  \quad w_{\rm{de},\Lambda}=-1. 
\end{aligned}
\end{equation}
Similar to EDE models, this specific interaction model modifies the dynamics of the universe around the matter-radiation equality era, due to the altered redshift dependency of relativistic relics as $\rho_{\nu} \propto a^{-4-\frac{4\alpha}{2\alpha+\sqrt{3}}}$ which is obtained from Eq.~\eqref{cont-nu}.

Before closing this section, we should remark that in the case of IDE models, we interpret the contributions arising only from the EMSF as a DE component whereas in the discussion of relativistic species, we, as in Ref.~\cite{Akarsu:2018aro}, also include the contribution coming from the usual material field together with EMSF in the DE. Besides the two approaches presented in this study, there is also another method exercised in the literature~\cite{Akarsu:2019ygx}: although a source in the dark sector nonminimally interacts with its EMSF, the authors of this work still assume that it behaves as in minimal interaction, viz., $\rho \propto a^{-3(1+w)}$ and $p= w \rho$, thereby, add the remaining terms in the altered evolution of the energy density and pressure into those of the DE as well.

 \subsection{EMSF as Hoyle-type Creation Field}
From another perspective, $T_{\mu\nu}^{\rm emsf}$ is reminiscent of the \textit{creation field} tensor introduced by Hoyle~\cite{Hoyle:1948zz} to modify the EFE, viz.,
\begin{equation}
\label{eq:cmunuhoyle}
C_{\mu\nu}=\kappa T_{\mu\nu}^{\rm emsf}.
\end{equation}
In this way, adhering to the perfect cosmological principle, which states that the Universe is homogeneous and isotropic in space as well as homogeneous in time~\cite{Bondi:1948qk}, Hoyle has achieved a steady-state model of the universe in the presence of dust whose energy density remains unchanged in an expanding universe due to a continuous creation of matter. The particular case  $C_{00} =0$ gives rise to Hoyle's steady-state universe model, namely, the time component of EMSF contributing to the energy density equation is arranged to vanish as follows:
\begin{equation}
\begin{aligned}
\label{createdT}
C_{00}=\kappa T_{00}^{\,\rm emsf}=\kappa f\,g_{00}-2 \kappa f_{\mathbf{T^2}} \theta_{00}=0.
\end{aligned}
\end{equation}
In the presence of the barotropic perfect fluid having constant EoS parameter $w$, we obtain
\begin{equation}
\begin{aligned}
f=\alpha'\left(\frac{T_{\mu\nu}T^{\mu\nu}}{1+3w^2}\right)^{\frac{1+3w^2}{2 (1+3w)(1+w)}}=\alpha'\rho^{\frac{1+3w^2}{(1+3w)(1+w)}
}, \quad 
\end{aligned}
\end{equation}
for the function $f$. Note that here $\alpha'$ is the integration constant and related to the model parameter $\alpha$ with $\alpha'=\alpha (1+3w^2)^{\frac{1+3w^2}{2 (1+3w)(1+w)}}$. Also, the EMSF contribution to the pressure equation is 
\begin{equation}
\begin{aligned}
 C_{ii}=\kappa T_{ii}^{\rm emsf}=\kappa fg_{ii}\,, 
\end{aligned}
\end{equation}
where $i=\{1,2,3\}$ denotes the spatial coordinates. Given the spatially flat FLRW spacetime metric stated above, the field equations therefore become
\begin{align}
\label{eq:fried}
3 H^2=&\kappa\rho,  \\
-2\dot{H}-3H^2=& \kappa w\rho+\kappa\alpha'\rho^{\frac{1+3w^2}{(1+3w)(1+w)}}.
\label{eq:pres}
\end{align}
As seen, the resulting source whose pressure has a linear term in energy density accompanied by a nonlinear function of it, and can be written as
\begin{align}
p_{\rm tot}=w \rho+\frac{\alpha'}{\rho^{-\frac{1+3w^2}{(1+3w)(1+w)}}},
\end{align}
mimics \textit{the modified generalized Chaplygin gas} (mGCG) \cite{Benaoum:2002zs,Bouhmadi-Lopez:2015oxa}. Yet the power of $\rho$ is $w$-dependent, as the interaction of each fluid with the EMSF is governed by the type of the fluid.

Note that dust (d) described by $w=0$ is indeed pressureless, but effectively gains pressure due to the interaction with EMSF. Namely, in this model, dust satisfies the following Friedmann equations 
\begin{align}
3 H^2=&\kappa\rho_{\rm d0}a^{-3(1+\alpha')},  \\
-2\dot{H}-3H^2=&\kappa \alpha'\rho_{\rm d0}a^{-3(1+\alpha')}.
\end{align}
Hence, it is reminiscent of a barotropic fluid with a constant EoS parameter equal to $\alpha'$. Hoyle's original steady-state universe model can be reproduced in the particular form $f(T_{\mu\nu}T^{\mu\nu})=\alpha'\sqrt{T_{\mu\nu}T^{\mu\nu}}$, the scale-independent EMSF, when $\alpha'=\alpha=-1$, for which $\rho_{\rm d}={\rm const}$. However, different from Hoyle's model, one can also achieve steady-state universe models in the presence of sources other than dust, as shown in Ref.~\cite{Akarsu:2023nyl} for the scale-independent case.

\section{Remarks and Future Perspectives}

In the gravity theories that incorporate a scalar formed from the usual Energy-Momentum Tensor (EMT) such as $g_{\mu\nu} T^{\mu\nu}$~\cite{Harko:2011kv},  $T_{\mu\nu}T^{\mu\nu}$~\cite{Katirci:2014sti,Roshan:2016mbt,Akarsu:2017ohj,Board:2017ign}, $R_{\mu\nu}T^{\mu\nu}$~\cite{Haghani:2013oma,Odintsov:2013iba} and $G_{\mu\nu} T^{\mu\nu}$~\cite{Asimakis:2022jel} in their actions, the second metric derivative of the matter Lagrangian density, $\frac{\partial^2 \mathcal{L}_{\rm m}}{\partial g^{\mu \nu} \partial g^{\sigma\epsilon}}$ arises as a new term, which does not exist in the literature to date, at the level of field equations. This term has been assumed to be zero for perfect fluids in the works considering these types of theories. A detailed investigation to provide a clear explanation for the vanishing of this term has revealed the proper meaning of a particular class of gravity models studied so far. In this study, we have shown that gravity models such as $f(\mathcal{L}_{\rm m})$~\cite{Harko:2010mv}, $f(g_{\mu\nu} T^{\mu\nu})$~\cite{Harko:2011kv} and  $f(T_{\mu\nu} T^{\mu\nu})$~\cite{Katirci:2014sti,Roshan:2016mbt,Akarsu:2017ohj,Board:2017ign} that modify the introduction of the material source in the usual EH action by adding only matter-related terms, namely, an arbitrary function of $\mathcal{L}_{\rm m}$ or those of scalars constructed from the usual EMT, $T_{\mu\nu}$, to the matter Lagrangian density $\mathcal{L}_{\rm m}$ cannot be considered as modified theories of gravity but rather are equivalent to General Relativity (GR) in the presence of nonminimally interacting sources. Thereby, assuming the second metric derivative of $\mathcal{L}_{\rm m}$ as zero turns out to be removing this term from the form of the interaction as a freedom. 
In fact, we see that defining the relation between the matter Lagrangian density and the EMT in such a way that $T_{\mu\nu}$ emerges on the right-hand side of the EFE of GR does not leave us room to alter the form in which the material fields appear in these equations. With the redefinition of the matter Lagrangian density as $\mathcal{L}_{\rm m}+f \rightarrow \mathcal{L}_{\rm m}^{\rm tot}$, we remain within the framework of GR while allowing for a nonminimal interaction between the usual material field $T_{\mu\nu}$ and its accompanying partner, $T_{\mu\nu}^{\rm mod}$, which is the modification field. At the level of field equations, this nonminimal interaction can be expressed as $\nabla^{\mu}T_{\mu\nu}=-Q_{\nu}=-\nabla^{\mu}T_{\mu\nu}^{\rm mod}$ where $Q_{\nu}$ is the interaction kernel that governs the rate of energy transfer. Therefore, the EoS of the fluid described by $T_{\mu\nu}^{\rm mod}$ and the form of the interaction described by $Q_{\nu}$ are intertwined in this type of models as both of them contain the function $f$ and the metric variation of the scalar ($\mathcal{L}_{\rm m}$, $g_{\mu\nu} T^{\mu\nu}$, $T_{\mu\nu} T^{\mu\nu}$) specifying the model. To elaborate on the discussion, we have focused on a particular type of these models known as the Energy-Momentum Squared Gravity (EMSG)~\cite{Katirci:2014sti,Roshan:2016mbt,Akarsu:2017ohj,Board:2017ign}. From the interaction perspective, in this model, the usual field is accompanied by \textit{Energy-Momentum Squared Field} (EMSF), $T_{\mu\nu}^{\rm emsf}$, determined from the arbitrary function $f(T_{\mu\nu} T^{\mu\nu})$. Then, we have given a brief discussion on the derivation of the EMT of the perfect fluid from the matter Lagrangian densities $\mathcal{L}_{\rm m}=p$ and $\mathcal{L}_{\rm m}=- \rho$ with $\rho$ and $p$ being the energy density and pressure of that fluid, respectively. We have explicitly stated under which assumptions the metric derivatives of $p$ and $\rho$ are derived in order to be used in the calculation of the perfect fluid EMT. Since these matter Lagrangian densities depend on not only the metric tensor but also other dynamical variables, we have derived all the related equations of motion satisfied in GR. It is clear that the violation of the aforementioned equations of motion due to the extra matter-related terms in the action allows the violation of local conservation of the perfect fluid EMT even if it is defined from the matter Lagrangian density, $\mathcal{L}_{\rm m}$. We have also demonstrated the proper calculation of $\frac{\partial^2 \mathcal{L}_{\rm m}}{\partial g^{\mu \nu} \partial g^{\sigma\epsilon}}$ where we utilize the sound speed, and therefore, the EoS of the barotropic fluid. In $\mathcal{L}_{\rm m}=p$ case, this term diverges in the presence of dust so removing it is a reasonable choice to be able to construct a completely viable cosmological/astrophysical model. However, omitting a term that arises from the variation of an action obviously renders that action invalid. In other words, gravity models such as $f(g_{\mu\nu} T^{\mu\nu})$~\cite{Harko:2011kv}, $f(T_{\mu\nu} T^{\mu\nu})$~\cite{Katirci:2014sti,Roshan:2016mbt,Akarsu:2017ohj,Board:2017ign}, $f(R_{\mu\nu} T^{\mu\nu})$~\cite{Haghani:2013oma,Odintsov:2013iba}, $f(G_{\mu\nu} T^{\mu\nu})$~\cite{Asimakis:2022jel} as well as their $f(R)$~\cite{Sotiriou:2008rp,DeFelice:2010aj} and teleparallel gravity~\cite{Bahamonde:2021gfp,Hohmann:2021ast} generalizations that set this term to zero for the perfect fluid work properly only at the level of field equations. Lastly, we have revisited the cosmological models in EMSG from nonminimal interaction perspective. We have investigated the possible effects of the EMSF interaction in the dark sector. It is possible to induce a dynamical DE component by constructing a model in which well-known sources such as baryons and photons have no EMSF partners while sources such as CDM and relativistic species nonminimally interact with their accompanying EMSFs. We have also shown that apart from giving rise to interacting DM-DE models, EMSF is indeed a Hoyle-type creation field. In the scale-independent form of EMSG described by $f(T_{\mu\nu} T^{\mu\nu})=\alpha \sqrt{T_{\mu\nu} T^{\mu\nu}}$~\cite{Akarsu:2018aro}, when $\alpha=-1$ we recover the Hoyle's original steady-state universe model~\cite{Akarsu:2023nyl}. Furthermore, the EMSF interaction extends the steady-state universe model to fluids other than dust, yet with a nonarbitrary, species-dependent parameter.

In this study, we have considered cosmological models in which only a single fluid nonminimally interacts with its accompanying EMSF while other fluids couple to the spacetime with their usual EMTs only. In the presence of more than one usual field, each field is accompanied by its own EMSF partner avoiding cross-terms of different usual fields~\cite{Akarsu:2018aro}. That is, in such a case, the field equations are written in the following form:
$G_{\mu\nu}=\sum_i T_{\mu\nu}^{(i)}+ \sum_i T_{\mu\nu}^{\rm{emsf} (i)}$ where a superscript $(i)$ represents the $i$th material field. To construct nonminimally interacting multi-fluid models, we have more options even though we are still in GR such that the usual material fields, say, ($T_{\mu\nu}^1, T_{\mu\nu}^2$) do not directly interact with their accompanying EMSFs ($T_{\mu\nu}^{\rm emsf1}, T_{\mu\nu}^{\rm emsf2}$) but interact nonminimally with each other, while a nonminimal interaction exits between the corresponding EMSFs as well, and interestingly, the form of one interaction determines that of the other. For instance, the scale-independent EMSF will exactly give the most common interaction kernel adjusted by hands of Barrow and Clifton in Ref.~\cite{Barrow:2006hia}, and is in the form of $\dot{\rho}_1+3H(w_1+1)\rho_1=\beta H\rho_1+\zeta H\rho_2$ and
$\dot{\rho}_2+3H(w_2+1)\rho_2=-\beta H\rho_1-\zeta H\rho_2$, where $\beta$ and $\zeta$ are positive constants. Different than the ad hoc choice of interaction, in this new interpretation of matter-type theories, $\beta$ and $\zeta$ are nonarbitrary, species($w$)-dependent constants, and interesting cosmological scenarios arising from an additional set of solutions appear for some specific values of $\alpha$~\cite{ABK}.

We would like to also add that although  the $\frac{\partial^2 \mathcal{L}_{\rm m}}{\partial g^{\mu \nu} \partial g^{\sigma\epsilon}}$ term can be removed owing to a freedom in determining the form of the new tensor $\theta_{\mu\nu}$ for EMSF,  one can still search for another form of interaction that is fully obtained from the variation of the matter Lagrangian density $\mathcal{L}_{\rm m}^{\rm tot}=\mathcal{L}_{\rm m}+f(T_{\mu\nu}T^{\mu\nu})$. For instance, in the choice of $\mathcal{L}_{\rm m}=-\rho$, one can have a well-defined Lagrangian formulation for such models.  Hence, the full field equations can be properly derived from the matter Lagrangian density $\mathcal{L}_{\rm m}^{\rm tot}=-\rho+f$, e.g., for the scalar $T_{\mu\nu} T^{\mu\nu}$, they are obtained from $\mathcal{L}_{\rm m}^{\rm tot}=-\rho+f(T_{\mu\nu} T^{\mu\nu})$. However, the resulting models are still equivalent to GR with other forms of nonminimal interaction. Moreover, if we consider, as an alternative to the EMSF interaction, another $f(T_{\mu\nu} T^{\mu\nu})$ model  with the new tensor found in Eq.~\eqref{thetarho}, the corresponding Friedmann equations read
\begin{align}
3 H^2=&\kappa\rho -\kappa f, \\
-2\dot{H}-3H^2=& \kappa p +\kappa f-2 \kappa f_{\mathbf{T^2}} p\,(\rho+p) \left(1+3 \frac{\partial p}{\partial \rho}\right),
\end{align}
and we see that the energy density equation is contributed by only the function $f$. Note that Hoyle's steady-state universe model cannot be achieved in this interaction, but it is possible to construct IDE models.

Before closing the paper, we would like to mention that we aim at highlighting the cosmological consequences of the EMSF interaction model in the work presented here, which can be seen as a phenomenological contribution to exploring the scope of possibilities. On the other hand, one may question the underlying microscopic physics of the EMSF interaction; in particular, whether there is a way of realizing such an interaction in the action within a particular field theoretical model that leads to the EMT~\cite{Pan:2020zza}. We know that there is a relationship between the quadratic EMSF described by the function $f(T_{\mu\nu}T^{\mu\nu}) \propto T_{\mu\nu}T^{\mu\nu}$ and loop quantum gravity \cite{Ashtekar:2006wn,Ashtekar:2011ni} as well as braneworld scenarios \cite{Sahni:2002dx,Brax:2003fv}, all of which add quadratic contributions of the material stresses' energy density to the Friedmann equation, hence, it would be interesting to look for a potential origin of the general form of EMSF interaction in a theory of fundamental physics and see whether such a relationship could be found.

\appendix*
\section{The scheme of the study}
\label{app:scheme}
\begin{figure}[ht!]
\begin{center}
\includegraphics[trim =0mm  0mm 0mm 0mm, clip, width=0.46\textwidth]{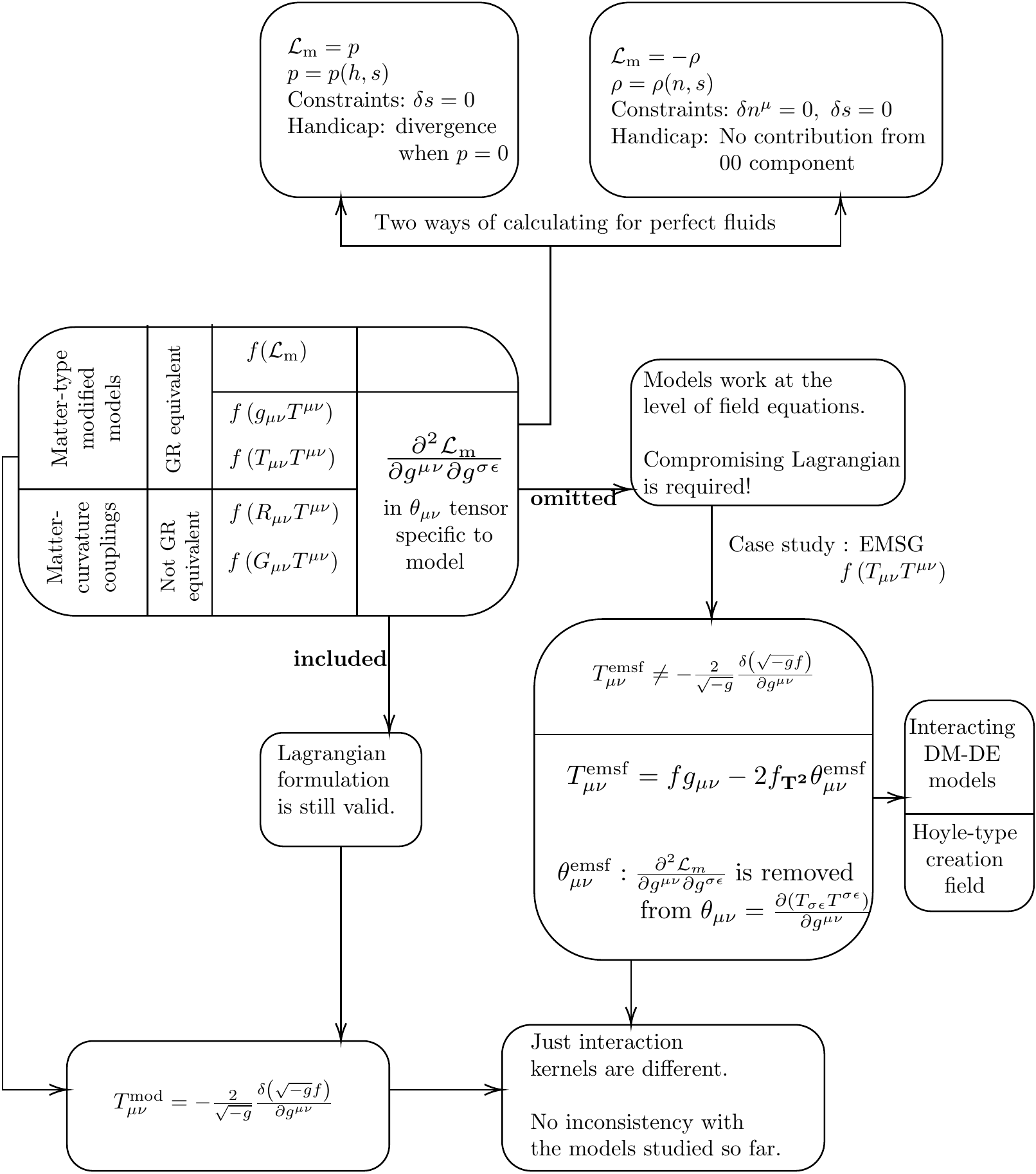}
\end{center}
\caption{The scheme of the study}
\label{fig:scheme}
\end{figure}

Fig.~\ref{fig:scheme} outlines the structure of our study, ensuring that each research question is addressed accordingly. It acts as a blueprint for the research presented in the sections that follow.

\newpage

\begin{acknowledgments}

\"{O}.A. acknowledges the support by the Turkish Academy of Sciences in the scheme of the Outstanding Young Scientist Award  (T\"{U}BA-GEB\.{I}P). \"{O}.A. and N.K. are supported in part by T\"{U}B\.{I}TAK grant 122F124. N.K. thanks Do\u gu\c s University for the financial support provided by the Scientific Research (BAP) project number 2021-22-D1-B01. M.B.L. is supported by the Basque Foundation of Science Ikerbasque and has
been financed by the Spanish project PID2020-114035GB-100 (MINECO/AEI/FEDER, UE). M.B.L. also
would like to acknowledge the financial support from the Basque government Grant No. IT1628-22
(Spain). N.M.U. was supported first by Boğazi\c ci University Research Fund Grant Number 18541P, and then by T\"{U}B\.{I}TAK grant 122F124 throughout this project.  E.N. and M.R. gratefully acknowledge the support by Ferdowsi University of Mashhad. This article is based upon work from COST Action CA21136 Addressing observational tensions in cosmology with systematics and fundamental physics (CosmoVerse) supported by COST (European Cooperation in Science and Technology). 
\end{acknowledgments}

\end{document}